\pdfoutput=1
\documentclass[aps,prb,twocolumn,superscriptaddress,amsmath,amssymb,showpacs]{revtex4-1}
\usepackage{graphicx}
\usepackage{physics}
\usepackage{siunitx}
\usepackage{amsmath}

\usepackage[utf8]{inputenc}

\newmuskip\pFqmuskip

\newcommand*\pFq[6][8]{%
  \begingroup % only local assignments
  \pFqmuskip=#1mu\relax
  \mathcode`\,=\string"8000
  \begingroup\lccode`\~=`\,
  \lowercase{\endgroup\let~}\pFqcomma
  {}_{#2}F_{#3}{\left(\genfrac..{0pt}{}{#4}{#5};#6\right)}%
  \endgroup
}
\newcommand{\pFqcomma}{\mskip\pFqmuskip}
\usepackage{cleveref}

\begin{document}

\title{Identical Spin Rotation Effect and Electron Spin Waves in Quantum Gas of Atomic Hydrogen}

\author{L. Lehtonen}
\author{O. Vainio}
\author{J. Ahokas}
\author{J. J{\"a}rvinen}
\author{S. Novotny}
\author{S. Sheludyakov}
\affiliation {Department of Physics and Astronomy, University of Turku, 20014 Turku, Finland}
\author{K.-A. Suominen}
\author{S. Vasiliev}
\email{servas@utu.fi}
\affiliation {Department of Physics and Astronomy, University of Turku, 20014 Turku, Finland}
\author{V. Khmelenko}
\author{D. M. Lee}
\affiliation {Institute for Quantum Science and Engineering,  Department of Physics and Astronomy, Texas A$\&$M University, College Station, TX, 77843, USA}

\date{\today}

\begin{abstract}
We present an experimental study of electron spin waves in atomic hydrogen gas compressed to high densities of $\sim 5 \times 10^{18}$ cm$^{-3}$ at
temperatures ranging from 0.26 to 0.6 K in strong magnetic field of 4.6 T. Hydrogen gas is in a quantum regime when the thermal de-Broglie wavelength is much larger than the \textit{s}-wave scattering length. In this regime the identical particle effects play major role in atomic collisions and lead to the Identical Spin Rotation effect (ISR). We observed a variety of spin wave modes caused by this effect with strong dependence on the magnetic potential caused by variations of the polarizing magnetic field. We demonstrate confinement of the ISR modes in the magnetic
potential and manipulate their properties by changing the spatial profile of magnetic field. We have found that at a high enough density of H gas the magnons accumulate in their ground state in the magnetic trap and exhibit long coherence, which has a profound effect on the electron spin resonance spectra. Such macroscopic accumulation of the ground state occurs at a certain critical density of hydrogen gas, where the chemical potential of the magnons becomes equal to the energy of their ground state in the trapping potential. 
\end{abstract}

%\pacs{67.63.Gh, 67.30.hj, 32.30.Dx, 32.70.Jz}

\maketitle

\section{introduction}
The concept of spin is one of the main features which distinguish between quantum and classical behaviour of matter. Interactions between particles having spin, obeying laws of quantum mechanics, may lead to a transport of spin variable in space and time, or propagation of the spin-perturbation in a wave-like manner. Spin waves in solid systems, e.g. ferromagnets, usually result in the exchange or dipolar interactions of electrons, which have sufficient overlap of their wavefunctions. Spin waves related to the nuclear spin of $^3$He represent a special case of this phenomenon in a degenerate Fermi liquid, where a large variety of quantum effects are observed. Quantization of the spin-wave oscillations leads to a description of the system as a gas of quasi-particles, or magnons. Such quantized waves or collective excitations of particles, similar to the normal atoms, may exhibit quantum phenomena, dependent on the type of statistics they obey. For Bose-type of (quasi)particles one of the most famous is the phenomenon of Bose-Einstein condensation (BEC).  Being first considered for the light quanta \cite{Bose_1924, Einstein_1924}, statistical attraction between identical bosons leads to the macroscopic occupation of their energy ground state, when their number exceeds a certain critical value for a given temperature. For cold atoms the phenomenon of BEC has been demonstrated in 1995 \cite{Anderson_1995, Davis_1995} and has been intensively studied during last decades. 

The main difficulty in the statistical treatment of the BEC of quasiparticles lie in the fact that, like for the light quanta of black body radiation, their chemical potential is zero in thermal equilibrium, and the quasiparticle number is provided by the external reservoir in amounts dictated by its temperature. Lowering the temperature decreases the occupation numbers of different quantum states including the ground state, and no criticality is reached with respect to T. Therefore, one has to consider a non-equilibrium case, when the excessive (over equilibrium) number is created by e.g. injecting quasiparticles with some external source. In such a case the chemical potential becomes non-zero, and the BEC regime can be attained with macroscopic population of the ground state. Such BEC-like behavior, or spontaneous coherence, had been predicted by Fr{\"o}hlich \cite{Frohlich_1968} in 1968 for systems of coupled oscillators and is often referred to as the Fr{\"o}hlich coherence. Recently these effects were observed for such quasiparticles as exciton polaritons \cite{Kasprzak_2006}, triplet states in magnetic insulators \cite{Ruegg_2003}, magnons in ferromagnets \cite{Demokritov_2006} and liquid $^3$He \cite{Autti_2012}, photons in a microcavity \cite{Klaers_2010}, and for magnons in a gas of spin-polarized atomic hydrogen (H$\downarrow$) \cite{Vainio2015}.

Spin waves in cold gases are fundamentally different from other systems because they are generated from quantum collisions of identical particles, basically during very short times when the particles approach each other to the distance comparable with their degree of delocalisation in space given by the de-Broglie wavelength. Exchange effects of the particles lead to the rotation of their spins around the total spin, called Identical Spin Rotation Effect (ISR), and to a wave-like propagation of the spin perturbation. The spin waves associated with ISR were first observed for the nuclear spins in the gas of $^3$He \cite{Nacher_1984} and atomic hydrogen \cite{Johnson_1984}, and recently for the cold gas of $^{87}$Rb \cite{McGuirk_2002} and for electron spins of atomic hydrogen \cite{Vainio2012}. Spin waves in atomic hydrogen represent a special case because they are studied in strong magnetic fields where the nuclear and electron spins are decoupled, and each can serve as the propagating quantity.

In this paper we present a description of our experimental study of the electron spin waves in the cold gas of atomic hydrogen. The work is done in a strong magnetic field of 4.6 T, over a temperature range of 0.26-0.6 K, when the gas is not yet degenerate, but already sufficiently cold for studies the effects of quantum collisions. Short descriptions of our experiments were given in two Letters \cite{Vainio2012, Vainio2015}. In \cite{Vainio2012} we have reported the possibility of generation of electron spin waves, guiding and trapping them with a magnetic potential. In the reference \cite{Vainio2015} an emergence of spontaneous coherence and BEC-like behaviour was observed. Here we will present a more detailed description of the experimental apparatus and technique, further analysis of the experimental results, and theoretical modelling of the magnons behaviour.

\section{background}

The nature of the ISR effect requires the gas to be in the quantum regime. The gas is considered to be quantum when the thermal de-Broglie wavelength $\Lambda_{th}$ substantially exceeds the scattering length for elastic collisions $a_s$. The quantum gas condition is not the same as that for the degenerate quantum gas. The latter situation occurs when $\Lambda_{th}$ exceeds the mean interatomic spacing $r$, and it is associated with drastic changes in the macroscopic properties of the gas; the phenomenon of BEC of atoms. Therefore, there exists a rather wide experimentally accessible range of densities $n \sim r^{-3}$ and temperatures where $r > \Lambda_{th} > a$, i.e. the gas is in the quantum regime but not yet degenerate. The rapid progress to BEC with alkali atoms has left this parameter region quite neglected. This range is especially large for H because of the small size of the atom ($a_{H} \approx 0.07$ $n$m).  In the quantum gas regime there appears a special type of collisions when identical particles exchange (rotate) their spins without changing their momentum. Such Identical Spin Rotation effect was predicted by Bashkin \cite{Bashkin_1981} and Lhuillier and Lalo{\"e} \cite{Lhuillier_1982_I}. Accumulating in multiple subsequent collisions ISR may lead to the propagation of spin waves. The efficiency of ISR depends on the ratio of the frequencies of spin changing collisions to the elastic collisions and is given by the so-called ISR quality factor $\mu^{*} \sim \Lambda_{th} / a_s$, which is large in the quantum gas regime. 
															
Soon after their theoretical prediction, spin waves of ISR type were observed for spin-polarized $^3$He gas in Paris \cite{Nacher_1984}, and for nuclear spins of atomic hydrogen gas at Cornell University \cite{Johnson_1984}. Progress in trapping and cooling of alkali vapours allowed observation of the ISR waves also for ultracold $^{87}$Rb \cite{McGuirk_2002}. In all above mentioned experiments the spin waves were associated with nuclear spins (H protons and $^3$He) or with the total spin of atoms in zero magnetic fields ($^{87}$Rb). Pure electron spin waves were much more difficult to observe, but eventually were observed for atomic hydrogen in strong magnetic fields \cite{Vainio2012}, and are considered in this work. 

In our considerations of the spin dynamics, we follow the treatment of Bouchaud and Lhuillier\cite{Bouchaud_1985}, which is specialized on the case of electron spin waves in atomic hydrogen gas.  In this case the ISR equation looks like
\begin{equation}
\pdv{S_+}{t} =  \left( \frac{D_0 }{1+i | \mu^{*}|} \laplacian - i \overbrace{|\gamma  B(r,z)|}^{ V(r)} \right) S_+ . \label{eq:isre}
\end{equation}

Here $S_+ = S_x+i S_y$ is the small transverse component of the $\vec{S} = (S_x, S_y, S_z)$ spin polarization (or magnetization) vector, such that $\quad S_x, S_y \ll S_z \approx S_0$, $B(r,z)$ is the external magnetic field in $z$-direction, $\gamma$ the gyromagnetic ratio of the spin involved, $D_0$ the axial spin diffusion coefficient, and $\mu^{*}$ the spin wave quality factor. Multiplying the equation by $i$ gives a Schrödinger-like equation with damping. The kinetic energy is, as would it be for a particle with effective mass $m^{*}\approx - \hbar \mu^{*}/2 D_{0} \epsilon$, and the potential term is defined by the Zeeman energy of the spin in the magnetic field  $B(r,z)$. This analogy allows treatment of the ISR magnons in a manner similar to that of real atoms in magnetic traps. We may use similar terminology to that used in the cold atoms field, and classify the magnons as high and low field seekers. The electron spin waves are attracted to the regions of strong magnetic field, making their behavior similar to the high field seeking atoms. Nuclear spin waves of H are low field seekers, with the effects of the potential for them being ~650 times weaker. However, there is one important difference in the trapping of the magnons.  It turns out to be possible to use the walls of the experimental chamber as a part of the trap. With a special choice of non-magnetic materials for the walls (e.g. liquid helium) one may realize nearly perfect reflective boundary conditions for the spin waves. Then, the combination of such walls with the magnetic field maximum allows building 3D trap for high or low- field seeking magnons. 

The ISR theory can be applied to any internal degree of freedom of colliding atoms, which is exchanged in the quantum collisions. It can be nuclear or electron spin \cite{Bouchaud_1985}, a pseudo-spin \cite{Lhuillier_1982_II}, or just some quantity, which makes atoms distinguishable. But it turns out that there is fundamental difference in the sense of the identical spin rotation, which depends on the type of the atoms and the spin statistics
involved. This is taken into account by the parameter  $\mu^{*}$  which  depends on the details of interatomic potential, and basically on the sign of the $s$-wave scattering length in the limit of cold collisions. The dynamics of ISR magnons for the above mentioned gases of $^3$He, $^{87}$Rb, and nuclear and electron spin waves of hydrogen may be completely different and reveal new and interesting features.

\section{Experimental basics}
Identical spin rotation is a weak effect, and atoms need to make many hundreds of spin-changing collisions before the spin perturbations traverse the sample cell from one end to another. Since the effective mean free path for spin changing collisions can be estimated as  $\lambda_{SE}\sim 1 /(n \Lambda_{th}^2) \sim 10 \mu$m, at the gas density of $10^{16}$ cm$^{-3}$ and temperature of 0.3 K, one can estimate that for the sample cell size of 0.5 mm the atom makes on the average 50 spin-changing collisions during its travel between the SC walls. Therefore, the  gas densities substantially higher than that are required to observe ISR spin wave modes. Such high densities cannot be reached with the standard trapping techniques used in BEC experiments with alkali vapours. In this work, we utilize the more conservative compression technique by reducing the volume of the sample surrounded by physical rather than magnetic walls.
\begin{figure*}
\includegraphics[width=1\textwidth]{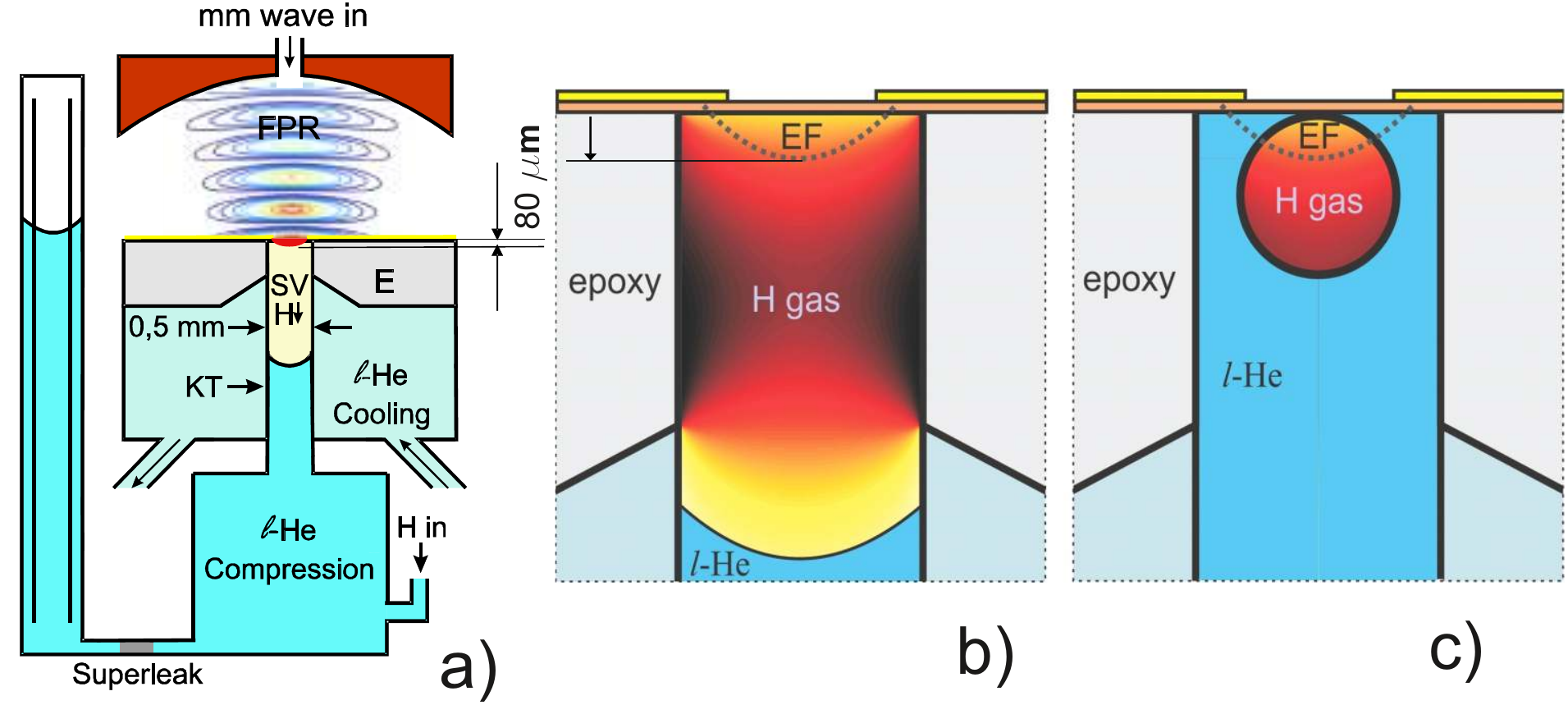}
\caption {a) Schematic representation of the sample cell and the compression technique. Helium piston is moved by the fountain pressure caused by the temperature difference between the left and right columns of the U-tube system separated by a superleak. Heating the left column transfers helium into it and opens the inlet of the H gas fill line. Then, ramping down the heating returns helium back to the right column with a variable cross section. Further compression pushes the gas to the top of the 0.5 mm Kapton tube (KT) with a typical $\sim 10^3$ fold reduction of its volume. Position of the helium piston is measured with submicrometer resolution using capacitance level meter in the left column. Evanescent field (EF) below the 0.4 mm iris in the lower mirror of of the 130 GHz Fabry-Perot resonator (FPR) is used for the excitation and detection of the ESR and spin waves. In order to remove recombination heat, H gas inside the thin ($\sim 3\mu$m) walled Kapton tube (KT) is actively cooled by a stream of superfluid helium (\textit{l}-He Cooling) flushing the outer walls of the capillary. b) Upper part of the compression region in case of the medium compression (I) in cylindrical geometry. c) Bubble stage of the strong compression (II). Colour map in the compression region illustrrates the remnant magetic field caused by small negative magnetization of the epoxy ring (E) around the upper part of the capillary. Black and dark red colour correspond to the stronger fields with a maximum of $\approx 0.3$G. }\label{fig:cell_1}
\end{figure*}

Atomic hydrogen gas is stabilized against recombination in a strong magnetic field of 4.6 T and temperatures $\sim$ 0.2-0.5 K. The electron and nuclear polarized state $\vert m_I, m_S \rangle = \vert -1/2, -1/2 \rangle$ is the only state which may be compressed to high densities of this work \cite{BlueBible}. The atoms are compressed with a fountain pump technique \cite{Vainio_2012} (see Fig. \ref{fig:cell_1} and Appendix A for details) using a piston of superfluid helium.  During the compression ramp the piston pushes the H gas to the top of the 0.5 mm diameter cylinder, reducing its volume by nearly three orders of magnitude and attaining densities above 10$^{18}$ cm$^{-3}$. At high densities considered in this work the number of atoms in the compressed gas decreases due to the process of the three-body recombination. We distinguish two different scenarios of the evolution of the gas sample after the compression ramp, according to which we will categorize experimental results presented in the following section.

I. Medium compression in \textbf{cylindrical geometry}: The compression ramp is stopped at 3-4 mm height of the gas cylinder and densities  $\sim 5 \times 10^{17}$ cm$^{-3}$. Decay of the gas density leads to the decrease of its pressure, and the helium piston moves up, approaching its value at zero density (Fig. 1 a).  The sample retains a cylindrical geometry with the height and gas density decreasing on a time scale of 10-60 minutes.

II. Strong compression ending with the \textbf{gas bubble}: Driving the sample height to smaller values leads to higher gas densities at the end of the compression ramp. Then, as in I, the height first decreases due to loss of the atoms in recombination. When the height becomes close to the tube diameter the sample geometry suddenly changes to that of the gas bubble. Now the surface tension of helium compresses the remaining H sample. The surface tension pressure is inversely proportional to the bubble radius, and for small enough bubbles is approximately equal to the gas pressure. The bubble rapidly shrinks and the gas density increases. Depending on the final height of the compression ramp the lifetime of the bubble ranges from 10 to 80 sec. Maximum densities $\approx 5\cdot 10^{18}$ cm$^{-3}$ are reached in the smallest bubbles of $\sim 20 \mu$m diameter which are still possible to detect with our technique. 

The combination of a magnetic field maximum with the helium covered wall can form a potential well where the spin wave modes described by the equation (1) can be excited and trapped. Therefore, the spatial distribution of the static magnetic field exerts a strong influence on the shape of the ESR lines as well as on the position, spacing and behaviour of their modulations resulting from the spin waves. 

Despite the special precaution for the choice of non-magnetic materials surrounding the compression region, we were not able to avoid tiny residual magnetic fields, which originated mainly from the Stycast 1266 epoxy ring forming the vacuum seal in the upper part of the compression tube. The weak diamagnetic properties of the epoxy in strong magnetic field (magnetization $M\approx -0.8$ G in the field of 46 kG) created a saddle profile of magnetic field with a maximum at the side-wall of the tube 0.5 mm below the top of the cylinder. It turned out that in the case of the medium compression (I) the magnetic potential associated with this epoxy field plays very important role in the dynamics of the spin waves. 

The field profile created by the magnetized epoxy ring cannot be changed and persisted in all experiments. We calculated it numerically using the known geometry of the ring, and using the epoxy magnetization as a fitting parameter to match the behaviour of the trapped magnon peak (see Appendix A). 
In addition to the permanent epoxy profile, we are able to add a linear axial gradient of magnetic field up to $\approx40$ $G/c$m using a system of gradient coils. In such strong gradients the inhomogeneities due to epoxy can be safely neglected in the analysis of the spectra.

In our data analysis below we distinguish three distinct cases with respect to the gradients and shape of magnetic potential : \textbf{large negative/positive gradient}, for which the magnetic field magnitude increases/decreases downwards, away from the rf excitation region (EF) located at the top of the cylinder with the compressed gas, and \textbf{saddle potential}, which is the magnetic field profile without any applied gradients.

We used electron spin resonance at 130 GHz for the excitation and detection of the electron spin waves. At the high densities used in this work it is not possible to use conventional techniques where the sample is located in a high quality resonator. The losses of microwave power at resonance are so large, that the response of the system cavity-sample becomes non-linear, and small variations of the absorbed power due to spin waves cannot be resolved. The 2.3 mm wavelength of the rf excitation is also too large for the selective excitation of spin-wave modes, which, as we will show below, have characterisic wavelengths of the order of several tens of $\mu$m. These two problems are solved by placing the compressed gas into the evanescent tail of the rf field (EF) leaking through the 0.4 mm diameter iris in the 0.5 $\mu$m thick planar mirror of the Fabry-Perot resonator. The field decreases exponentially below the top of the compression region with the characteristic length of 80 $\mu$m.

ESR spectra are detected using a heterodyne cryogenic ESR spectrometer \cite{Vasilyev_2004}. The output signal of the spectrometer provides real (absorption) and imaginary (dispersion) parts of the complex reflection coefficient of the cavity as a function of magnetic field or frequency. In this way we distinguish two modes of operation: continuous wave (CW) and pulsed mode.

In the CW mode the rf at the fixed frequency $\omega_{ex}=2\pi f_{ex}$ and very low ($\leq 1$ nW) power is fed into the Fabry-Perot resonator. The static magnetic field $B_0$ is swept through the resonance value for this frequency $\omega_{ex}= \gamma_e B_0$. This is done with a separate sweep coil applying a small ramp-like offset field $\Delta B_{sweep}(t)$. The absorption and dispersion signals are recorded as functions of $\Delta B_{sweep}$.

In the pulsed ESR mode the static magnetic field is fixed at some value near the resonance, and the excitation pulse is applied with a fixed excitation frequency and a given duration time $\tau_p$. The Free Induction Decay (FID) after the pulse is recorded and averaged for several thousand similar excitation-detection cycles. Then the Fourier transform (FT) is taken, which gives the real and imaginary components of the transverse magnetization as a function of frequency. The pulse duration defines the excitation width in the frequency domain $\delta f\approx 1/\tau_p$. If the width is larger than the CW ESR linewidth, the entire line is excited  and the Fourier transform of the FID provides a line-shape
similar to that seen with CW ESR. The possibility of exciting a small region of the inhomogeneously broadened line using narrow selective pulses leads to a significant advantage for the pulsed technique. This technique is substantially faster and allows acquisition times for one spectrum well below 1 sec. This is very important for detection of fast processes which occur in the experiments with bubbles.

Since the offset field $\Delta B_{sweep}$ is added to the inhomogeneous field inside the sample, the regions of stronger local field would appear on the left side of the CW ESR spectrum. The spectrum would be seen as the mirror image of the pulsed spectrum, in which the regions of stronger field oscillate at higher frequency. In order to avoid confusion and make straightforward comparisons of the spectra obtained by these two techniques we will present below the CW absorption spectra as a function of the local static field $B_0$, i.e. mirror images of the absorption function of the $\Delta B_{sweep}$.

During the compression ramp and following evolution of the sample we collect the  data of the liquid helium level meter reading and ESR spectra of the compressed gas as a function of time. The temperature of liquid helium around the compressed gas, the position of helium piston in the end of the compression ramp, and the direction and strength of the linear magnetic field gradients are the main parameters which we vary from one compression-decay cycle to another. Using the level meter readings we can calculate the volume and the density of the compressed gas. These two parameters are coupled and change simultaneously during each experimental cycle.
    
\section{Results}
\subsection{Compressions in cylindrical geometry}
\subsection*{CW ESR spectra}
\begin{figure}
\includegraphics[width=1\columnwidth]{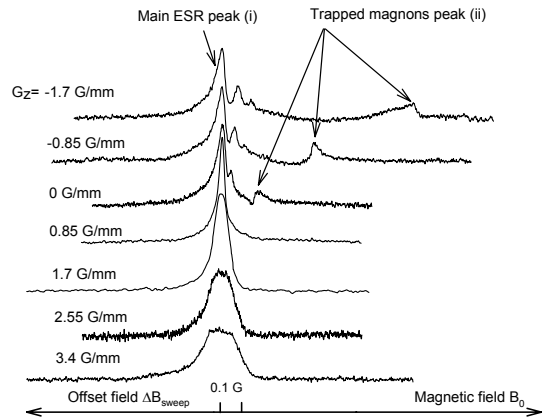}
\caption{ESR absorption spectra in various axial gradients of the static magnetic field. The gas density is $\approx8\cdot10^{16}$ cm$^{-3}$.}
\label{fig:Peaks in grad}
\end{figure}

First we analyse the behaviour of ESR spectra as a function of the axial gradients of magnetic field. In Fig. \ref{fig:Peaks in grad} we presented spectra at fairly small densities $n_H \approx 8\cdot 10^{16}$ cm$^{-3}$, where we observe sufficiently strong ESR signals. 

The spectra feature the main ESR peak (i) associated with the absorption by the H gas in the EF region. The peak is modulated by the spin-wave oscillations already at this density, which is clearly seen in spectra at zero and negative gradients. Position of the main ESR peak does not change after applying the axial gradient since the gradient coil system is well centered within the EF region. An extra peak in the high field side (ii) appears at zero and negative gradients, with the position dependent on the gradient strength. The change in position as a function of the gradient allows us to conclude that this peak originates from the region of space $\approx 0.5$ mm below the EF, i.e. from the maximum of the saddle potential created by the epoxy ring.  The distance between the (i) and (ii) peaks in zero gradient provides the value of the field difference $\Delta B \approx 0.25$ G between the maximum of the saddle potential and the EF. 

Applying small positive gradient leads to a partial cancellation of the natural inhomogeneity created by the epoxy ring. The ESR line narrows, and reaches minimum width at $\nabla B_0 \approx 1.7$ G/mm.  Furthermore, no extra peak appears on the left from the main ESR line at larger positive gradients. This proves that the peak (ii) is associated with the magnetic field maximum, and is not caused by a possible second maximum of the rf field located below EF. In the latter case the (ii) peak would appear as a mirror image on the left side of the main ESR peak upon reversal of the gradient. Superposition of the positive gradient with saddle profile shifts the field maximum up until it reaches the top of the cylinder. Then the peak (ii) merges with the ESR  peak (i). This behaviour of the peak (ii) suggests an interpretation in terms of spin wave modes trapped in the minimum of the saddle potential \cite{Vainio2012}. It is remarkable that we can observe these modes in the region where the excitation rf field is substantially smaller than in the EF. 

\begin{figure*}
\includegraphics[width=1\textwidth]{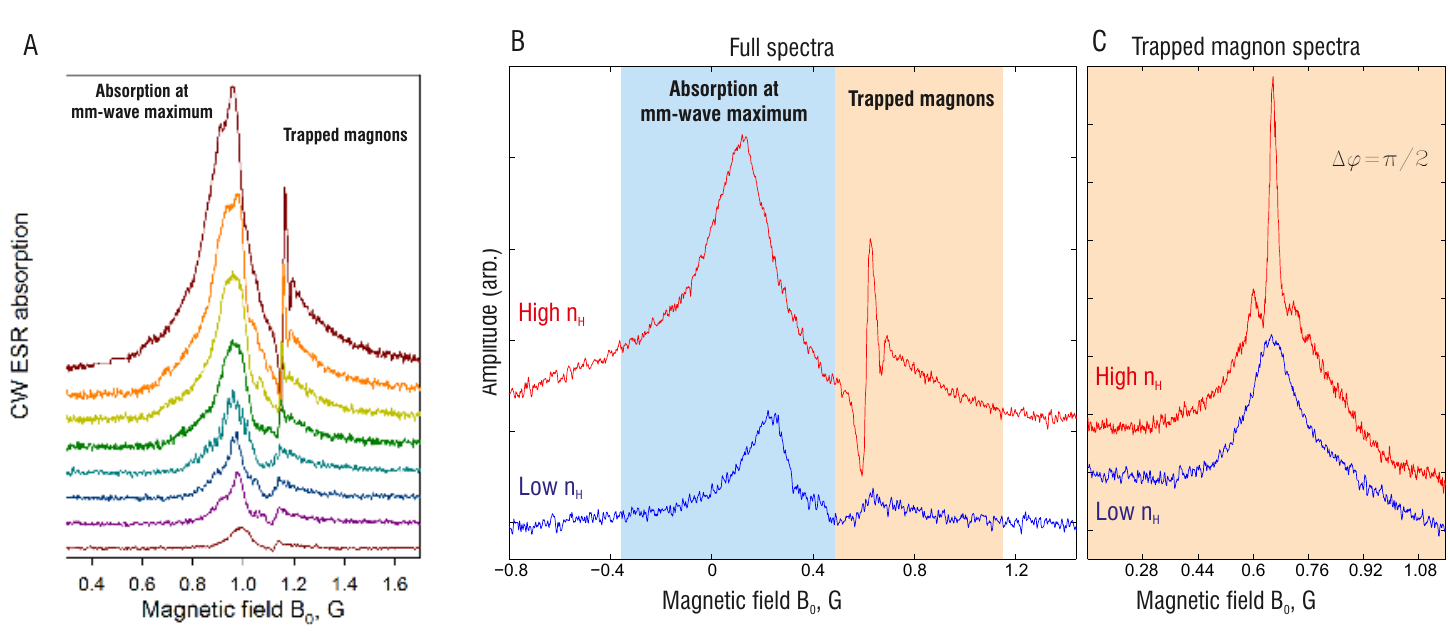}
\caption{A. Evolution of the ESR absorption spectra in the cylindrical geometry and natural gradient of magnetic field after the end of the compression ramp. Density decreases from top to bottom. B. Comparison of the spectra at low density $n_H\approx 5\cdot10^{16}$ cm$^{-3}$ (lower trace), and largest density $n_H \approx 4\cdot10^{17}$ cm$^{-3}$ reached just after the end of the compression ramp. C. Zoom into the trapped magnons peak for the same densities. The phase of the ESR signal is changed by $\pi /2$.}
\label{fig:Nat grad spectra BEC}
\end{figure*}

Next we look into the dependence of the ESR spectra on the H gas density.  In Fig. \ref{fig:Nat grad spectra BEC} we present the evolution of the ESR absorption in the saddle profile of magnetic field, with no extra gradients applied. One can see that at a certain high density, $\sim10^{17}$ cm$^{-3}$ a sharp and narrow feature appears in the trapped magnon peak. The phase of this feature differs by $\approx \pi /2$  (Fig. \ref{fig:Nat grad spectra BEC} B). Changing the phase of the detection by this amount, we reproduce the true lineshape of the trapped magnon peak in Fig. \ref{fig:Nat grad spectra BEC} C.  
The difference of $\approx \pi/2$ of the peak (ii) from the main ESR peak also indicates that the region in space where this oscillation occurs is located at $\approx$ quarter wavelength ($\lambda /4 \approx 0.5$ mm) from the EF region. The mm-wave needs to travel twice this distance and to acquire such a phase shift. This region coincides with the the position of the saddle magnetic trap.

\subsection*{Pulsed ESR spectra}
In order to investigate further the spectra in the natural gradient we have used the pulsed mode of detection of ESR spectra. As we have already mentioned, we can tune the spectral width of the excitation by adjusting the pulse duration in the time domain. Using short enough pulses, with width $\leq0.5 \mu$s we can cover the region $\geq 1$ G of the CW spectrum. In this case all atoms in this region are excited simultaneously and oscillate at the frequency given by the static magnetic field at their location. The Fourier transform of the FID evaluates the strength of these oscillations as a function of frequency. The spectral line has the shape nearly identical to the CW line, plotted as a function of frequency $2\pi f=\gamma_e B_0$. Using narrow pulses shaped with the gaussian or $sinc$ function we can excite selectively parts of the inhomogeneously broadened ESR line corresponding to certain regions of space. This is done by tuning the offset field to a desired position in the ESR line. The atoms from the excited region may then redistribute themselves in the compression volume and provide oscillating signals at all frequencies. If we consider ballistic flight in the molecular regime, the flight time from the EF to the minimum of the saddle potential is $\tau_{bal}\approx 6 \mu$s. At densities of $10^{17}$ cm$^{-3}$ the excited atoms distribute via diffusive motion, and the characteristic time to travel from the EF region to the saddle trap center is $\tau_{dif}\sim 200 \mu$s. This is substantially longer than the recombination time of the excited atoms. We recall that due to the tilted electron spins these atoms have much higher recombination rates than the atoms with fully polarized electron spins. Therefore, we should not expect any substantial spread and broadening of the ESR signal after selective pulses unless some other spin transport channels are involved.

\begin{figure}
\includegraphics[width=1\columnwidth]{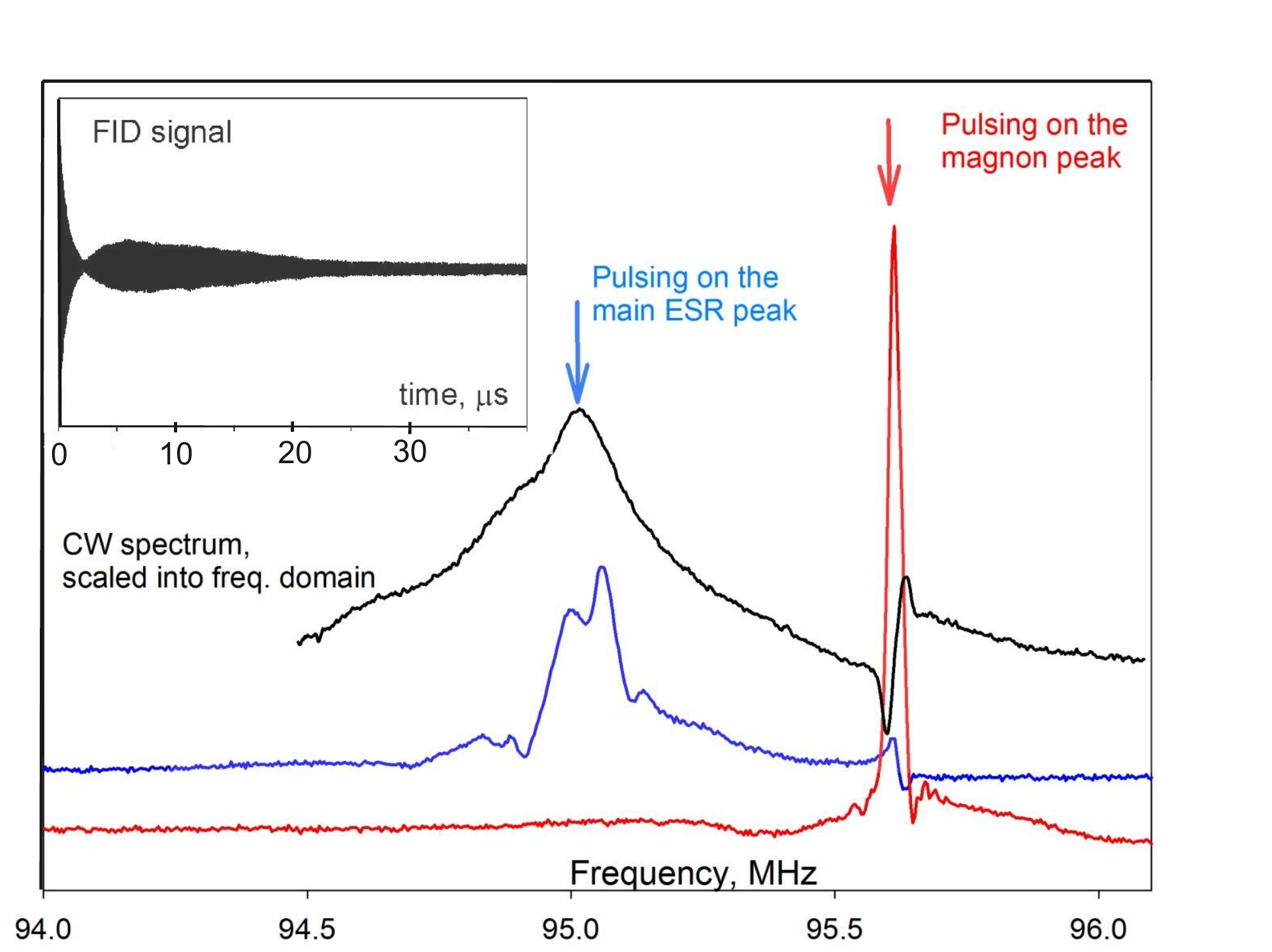}
\caption{Pulsed ESR spectra using 6$\mu$s long selective pulses of gaussian shape. This corresponds to the $\approx 100$ kHz spectral width of the excitation pulse. Black upper trace is the full ESR lineshape obtained by the CW method and plotted for comparison as a function of the excitation frequency $2\pi f= \gamma_e B_0$. Lower blue trace is the FT spectrum of the pulsed ESR when the magnetic field is tuned to the center of the main ESR peak. Middle red trace is recorded when the excitation is tuned in resonance with the center of the magnons trap. The pulsed spectra are recorded at fixed frequency, and static magnetic field is tuned to the desired region of the CW ESR line. The inset shows the free induction decay signal in the time domain recorded when the magnetic field is tuned to the trapped magnons peak (red trace)}
\label{fig:Spin transport}
\end{figure}

In Fig. \ref{fig:Spin transport} we present results of two pulsed ESR spectra when the excitation frequency is tuned to: a) the center of the main ESR peak, and b) the center of the saddle trap for magnons.
As expected, we do not observe any oscillation in the EF region when the atoms are excited in the saddle trap. We see only a very strong and narrow peak corresponding to the trap center. The situation is different when we tune the pulse to the main ESR peak. In addition to the broad signal from this region we observe a sharp and narrow peak from the trap. As in the CW experiment (Fig. \ref{fig:Nat grad spectra BEC}) its phase appears to be shifted by $\pi /2$ from the main ESR signal. 

\begin{figure}
\includegraphics[width=0.8\columnwidth]{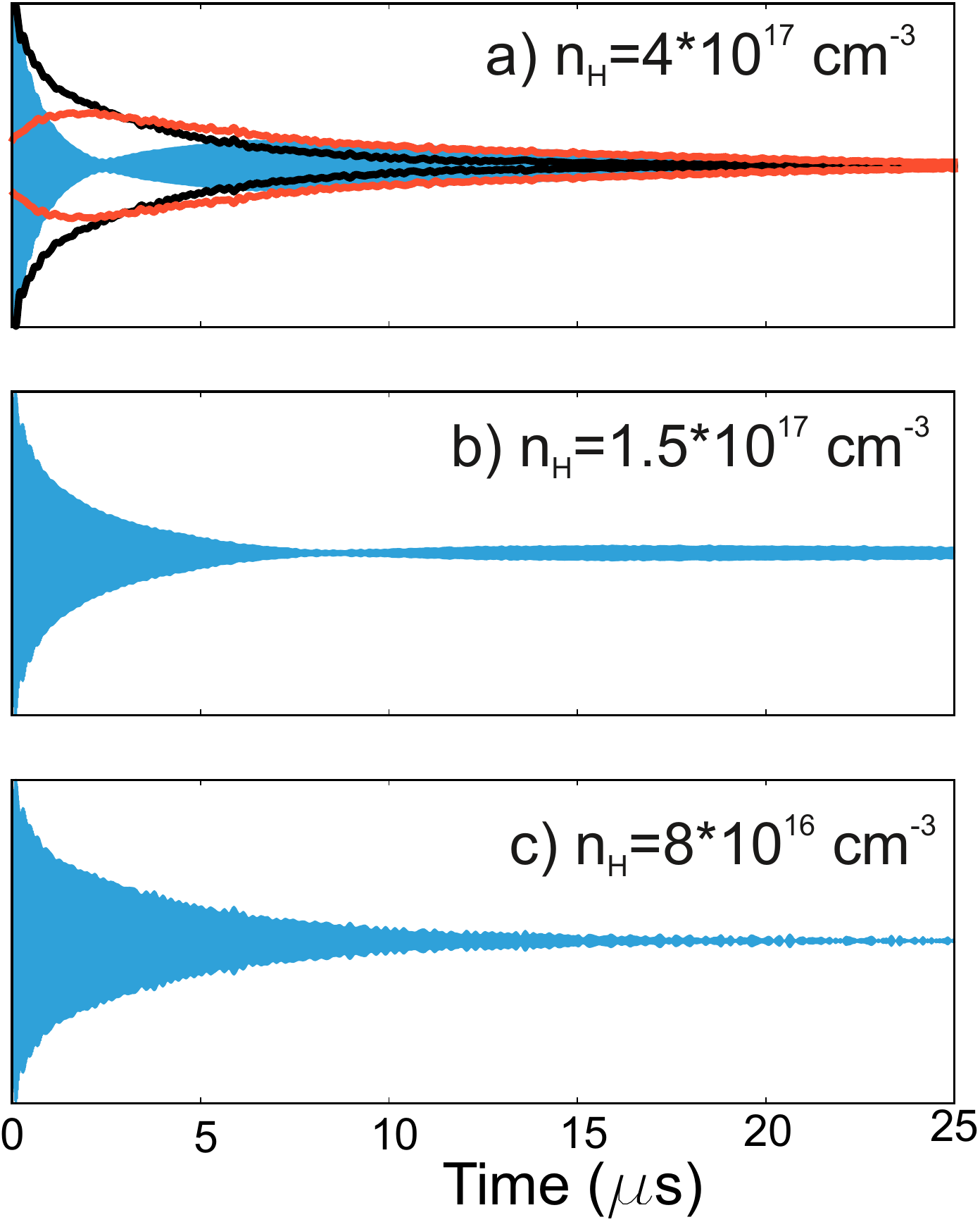}
\caption{Free induction decays recorded at different densities: well below the critical density c), well above a) and approximately at the critical density b) for the appearance of the sharp feature in the trapped magnons peak of the ESR spectrum. The selective ESR excitation pulses of $\tau_p =1 \mu$s duration are frequency centered with the trapped magnon peak. For the sake of comparison the signals amplitude is scaled to the same starting amplitude. Solid lines in a) are envelopes of the FID envelope deconvolution as described in the text: black line is the rapidly decaying oscillation taken from the FID in c), red line is the growing and then slowly decaying component remaining after subtraction.}
\label{fig:FID vs density}
\end{figure}

The time-domain FID data after the pulses centered on the trapped magnon peak for various H gas densities are presented in Fig. \ref{fig:FID vs density}. One can see that at low density (bottom trace) the free induction signals decay exponentially with a rather short time constant $T_2\sim 3\mu$s. The shape of the FID changes at a critical density of $n_H\approx1.3\cdot10^{17}$ cm$^{-3}$, the same value for which the sharp feature also emerges from the magnon peak in the CW spectrum. At high density the spin oscillations retain the rapidly decaying part in the beginning, but then the signal recovers and persists for several tens of $\mu$s. We note that the inhomogeneity of magnetic field in the saddle trap region would correspond to a pulsed ESR signal broadening of $\sim 1$ MHz. This is substantially larger than the observed $\leq 20$ kHz width of the strong peak originating from the saddle trap region. The characteristic growth time of the FID signal $\sim 5-6 \mu$s is substantially shorter than the diffusion time of H atoms from the EF region into the saddle trap.

Development of a node in the FID signals at high gas density may be explained by an interference of two oscillating signals: a rapidly decaying component together with a second, first growing and then slowly decaying spin oscillation. It is natural to assume that the fast decaying signal is of the same origin as the one for low densities (bottom plot in Fig. \ref{fig:FID vs density}). Then, we may decompose the FID at high densities, subtracting the fast decaying part from it. The remaining component indeed shows a growing part in the beginning and then slow decay. Envelopes of both components are presented as solid lines in the upper plot of Fig. \ref{fig:FID vs density}. Making separate Fourier transforms of the beginning part of the FID before the node and the long lasting oscillation after, we found that the frequencies of the oscillations may be slightly different. The difference depends on the width of the excitation pulse and its frequency tuning within the ESR spectrum.

We note that mathematically there is another possibility to reproduce the FID with a node as seen in the top plot of Fig. \ref{fig:FID vs density}. It can be obtained as a result of destructive interference of two decaying components oscillating at the same frequency and shifted in phase (e.g. by $\pi/2$ or $\pi$). However, we cannot find any physical reasons for a such situation. The superposition of signals at slightly different frequencies is expected for the inhomogeneously broadened ESR line. But the oscillations originating from the same region of space should also have the same phase. Therefore, we consider unlikely the possibility of such destructive interference.  

\subsection{Bubbles}
\begin{figure}
\includegraphics[width=1\columnwidth]{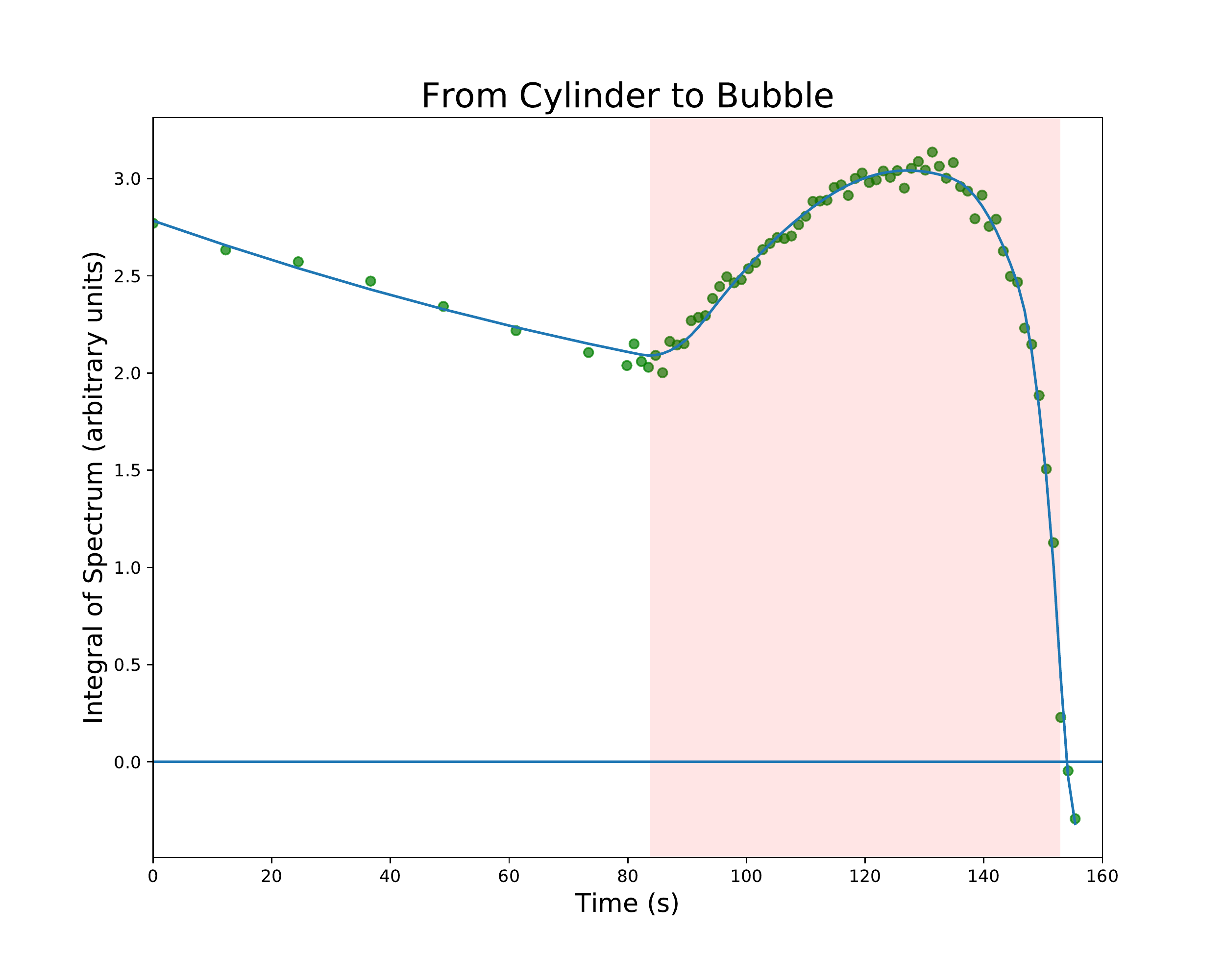}
\caption{Plot of the ESR line integrals at the stage of the transfer of the compressed gas from cylindrical geometry to the bubble.}
\label{fig:Cylinder-to-Bubble}
\end{figure}

The transition from the cylindrical to the bubble geometry occurs in sufficiently strong compressions when the height of the cylinder becomes comparable to its diameter. This event is clearly visible in evolution of the ESR spectra. In the cylindrical geometry the density of the gas decreases due to recombination. The integrals of the ESR line which are proportional to the number of H atoms in the EF region also decrease. Once the bubble is formed, the surface tension compresses the gas and its density starts to increase. This change is clearly seen in the plot of the integrals of the ESR line as a function of time (fig. \ref{fig:Cylinder-to-Bubble}). A similar kink is observed in the level meter readings, which provide the data on the volume of the gas sample. The growth of the integrals of the ESR line soon changes to a decrease. This happens when the bubble size becomes smaller than the size of the EF region. Now the decrease of the bubble volume and decrease of the number of H atoms seen by the ESR dominates over the growth of density.

\begin{figure*}
\includegraphics[width=0.9\textwidth]{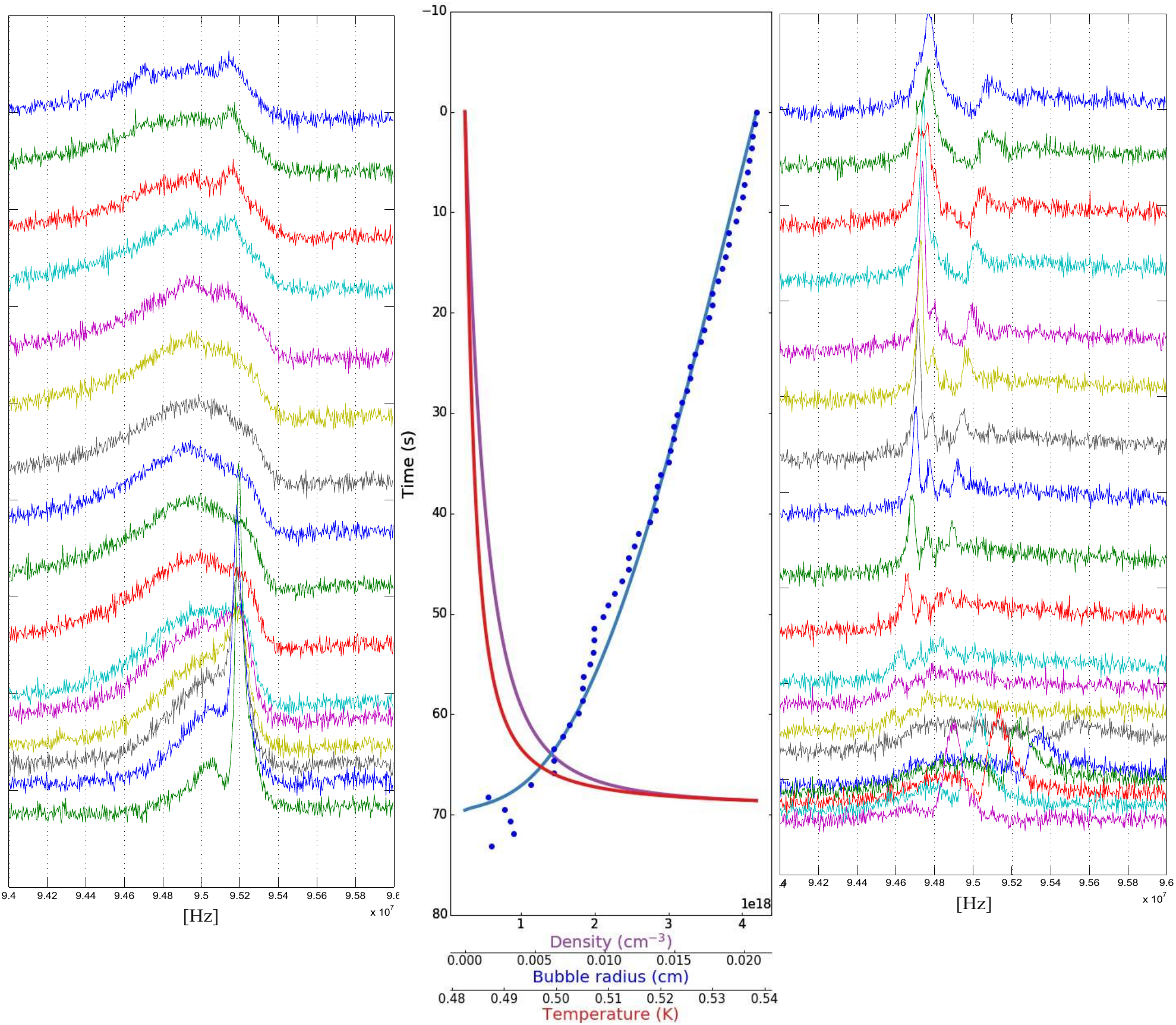}
\caption{Evolution of the ESR spectra and main parameters of the H gas in the bubble stage. Left/right panels present ESR spectra in large positive/negative gradients of static magnetic field $\triangledown B_0 = \pm 30$ G/cm. The center panel provides a plot of the main parameters of the gas bubble obtained in the simulation fitted to the ESR and levelmeter data (blue dots) in the initial part of the bubble decay. Diameter, density and temperature of the gas in bubble are plotted with blue, purple, and red colours accordingly. The time scale goes from top to bottom, and the ESR spectra are shifted vertically in line with the time scale.}
\label{fig:Bubble evolution}
\end{figure*}

Determination of the main parameters of the gas in the bubble can be reliably done only at the initial stage, when the gas density is extracted from the ESR line integrals, and the level meter data provide the volume of the bubble. The bubble size can be determined by the level meter down to the smallest diameter of $\approx 50 \mu$m. Once the bubble size gets smaller than the EF, determination of the gas density from the ESR integrals becomes very complicated. In order to evaluate the properties of the gas sample during further evolution we use numerical simulation of the bubble decay which is fitted to the ESR and levelmeter data in the beginning part of the decay. The simulation is based on the approach used in similar work with H gas bubbles at high density \cite{Tommila1987}. We solve a set of the coupled kinetic equations for each of the four hyperfine states which account for different recombination and relaxation mechanisms. The recombination and relaxation rate constants are taken from previous numerous experimental studies of H gas in strong magnetic field \citep{BlueBible}. Recombination results in heating and loss of atomic species. Heat transfer to the helium bath at the boundary of the bubble establish a temperature profile in the gas, which in turn affects the reaction rates in different parts of the gas. Heating and sample losses also change the volume to counter the bubble's surface tension $\frac{2\sigma(T)}{r}$. Results of the simulation are shown in the center panel of Fig. \ref{fig:Bubble evolution}. The density and temperature in the bubble rise sharply during the final moments of the bubble's existence.

ESR spectra shown in Fig. \ref{fig:Bubble evolution} exhibit one common feature: the strong and sharp peak grows rapidly at the end of the bubble life, starting at the critical density of $\approx 1.3 \cdot 10^{18}$ cm$^{-3}$. This is most clearly seen in data on the left panel recorded in a large positive gradient of magnetic field. In this case the peak position does not change in time. In contrast,  in the negative gradient the peak moves to lower frequencies. Such peak behaviour is consistent with the changes in the location of the magnetic field maximum inside the bubble. In a positive gradient the field maximum is at the top of the bubble and does not depend on the bubble diameter. In the negative gradient the field maximum appears at the bottom, which moves up when the bubble shrinks. The peak height is larger in the positive gradient since it has stronger overlap with the maximum of the excitation field. The sharp peak in the bubble occurs at about same density $n_H\approx 8\cdot 10^{18}$ cm$^{-3}$ for gradients of both signs, which is approximately an order of magnitude larger than in the cylindrical geometry.

\begin{figure}
\includegraphics[width=0.45\textwidth]{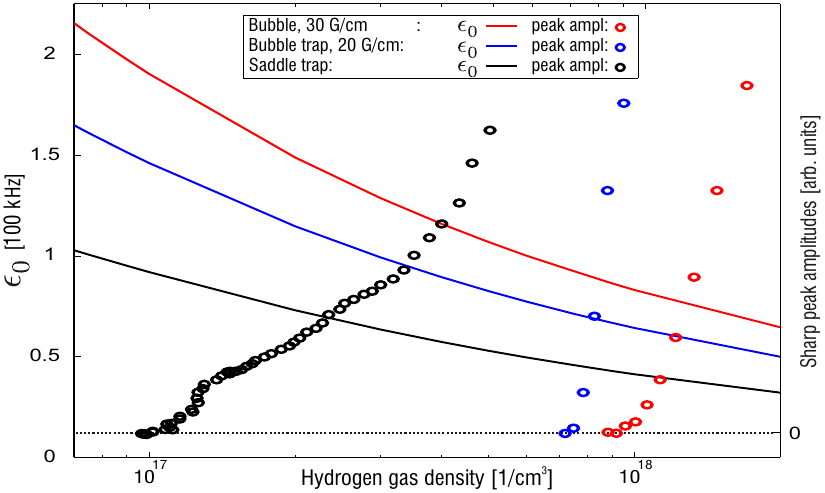}
\caption{Amplitudes of the sharp peak originating from trapped magnons spectrum in different geometries and magnetic field gradients. Solid lines are the plots of the ground state energies for all three cases.}
\label{fig:Critical densities}
\end{figure}

The strength of the magnetic field gradient is a parameter which we may change, and therefore we performed bubble compression experiments in various gradients. In Fig. \ref{fig:Critical densities} we present the sharp peak amplitudes as a function of H gas density for the bubble compression at two values of the field gradient 20 and 30 G/cm, and for comparison, the same data in the cylindrical geometry in the natural field profile. One can see that the critical density is larger for bubbles in stronger gradients. 

\section{Analyses}
\subsection{Solutions of the ISR Equation in Cylindrical Geometry}
The ISR equation (\ref{eq:isre}) can be solved numerically using magnetic potentials based on a calculation of the actual magnetic field profiles created by the magnetized epoxy ring and linear axial gradients created by the external coils presented in Appendix A. The presence of the saddle minimum of the magnetic potential creates an interesting situation when the lowest spin wave modes first occur in this potential and do not depend on physical boundaries. With increasing mode numbers their wave functions expand beyond the magnetic wall and spread in the whole cylinder volume. This behaviour is confirmed in numerical calculations based on actual spatial profiles of magnetic field. 

It turns out that with certain approximations for the magnetic field function it is also possible to solve Eq. \ref{eq:isre}  analytically. Since the structures and types of the spin wave modes are better understood in the latter case, in this section we will present analytic solutions of the ISR equation (\ref{eq:isre}) using the  approximation for the saddle field of the form $\vec{B}(r, z) = \left(B_0 + B_r r^2 - B_z z^2 \right)\vu{e}_z$ with   $B_0 =  \SI{4.6}{\tesla}$, $B_r = \SI{24e-3}{ \tesla \per \square \centi \meter} $, and $B_z = \SI{6.0e-3}{\tesla \per \square \centi \meter}$.
The ISR equation has a separable solution $S_{+}(t,r,z,\theta)=e^{i\omega t}R(r)Z_{{p, o}}(z)e^{i k \theta}$ in terms of confluent hypergeometric functions:

\begin{align*}
R\left(r\right)&=	r^{k}e^{-\frac{r^{2}\sqrt{i-\mu}}{2\lambda_{r}^{2}}}\pFq{1}{1}{\frac{k+1}{2}+\Omega_{l}^{k}}{k+1}{\frac{r^{2}}{\lambda_{r}^{2}}\sqrt{i-\mu}} \\
Z_{e}\left(z\right)=&	e^{-\frac{z^{2}}{2\lambda_{z}^{2}}\sqrt{\mu-i}}\pFq{1}{1}{\Gamma_{e}+\frac{1}{4}}{\frac{1}{2}}{\frac{z^{2}}{\lambda_{z}^{2}}\sqrt{\mu-i}} \\
Z_{o}\left(z\right)=&	ze^{-\frac{z^{2}}{2\lambda_{z}^{2}}\sqrt{\mu-i}}\pFq{1}{1}{\Gamma_{o}+\frac{3}{4}}{\frac{3}{2}}{\frac{z^{2}}{\lambda_{z}^{2}}\sqrt{\mu-i}}
\end{align*}
 
In the axial direction there are both even and odd solutions with $z=0$ in the middle of the cylinder. The complete solution can be characterized by a set of mode numbers $(k, l, m, p)$, where $k$ specifies the azimuthal mode, $l$ the radial mode, $m$ the axial mode, and $p$ whether the solution is even or odd. 

Since there is no spin flow into the walls of the compression cylinder, we use Neumann boundary condition $\eval{\pdv{S_+}{\vec{n}}}_{\partial V}=0$; in our case, we've chosen these to be at $z=\pm \SI{0.5}{\milli \meter}$ (for a total length of \SI{1.0}{\milli\meter}) and $r=\SI{0.25}{\milli \meter}$.  This gives three characteristic equations, one in the radial direction and two in the axial, for the even and odd solutions (see Appendix B for details). These equations can be numerically solved for $\Omega_l, \Gamma_m^p$ to give the modes and the complex eigenfrequency $\omega$. 
\begin{align*}
\omega = & \frac{D_{0}}{ \left(-\mu +i\right)^{\frac{1}{2}}} \\ 
\times &  \left(
	\frac{\left(-\mu +i\right)^{\frac{1}{2}}}{\lambda_{0}^{2}}+4\left[ \frac{ i\Gamma_{m}^p}{\lambda_{z}^{2}} -\frac{ \Omega_l^k }{\lambda_{r}^{2}} \right]
	\right) \\
\lambda_{r,z} = & \left(\frac{D_0}{B_{r,z} |\gamma|}\right)^{\frac{1}{4}}, \, \lambda_0 =  \left(\frac{ D_0}{B_0 |\gamma|}\right)^{\frac{1}{2}}
\end{align*}

\begin{figure*}
\includegraphics[width=0.9\textwidth]{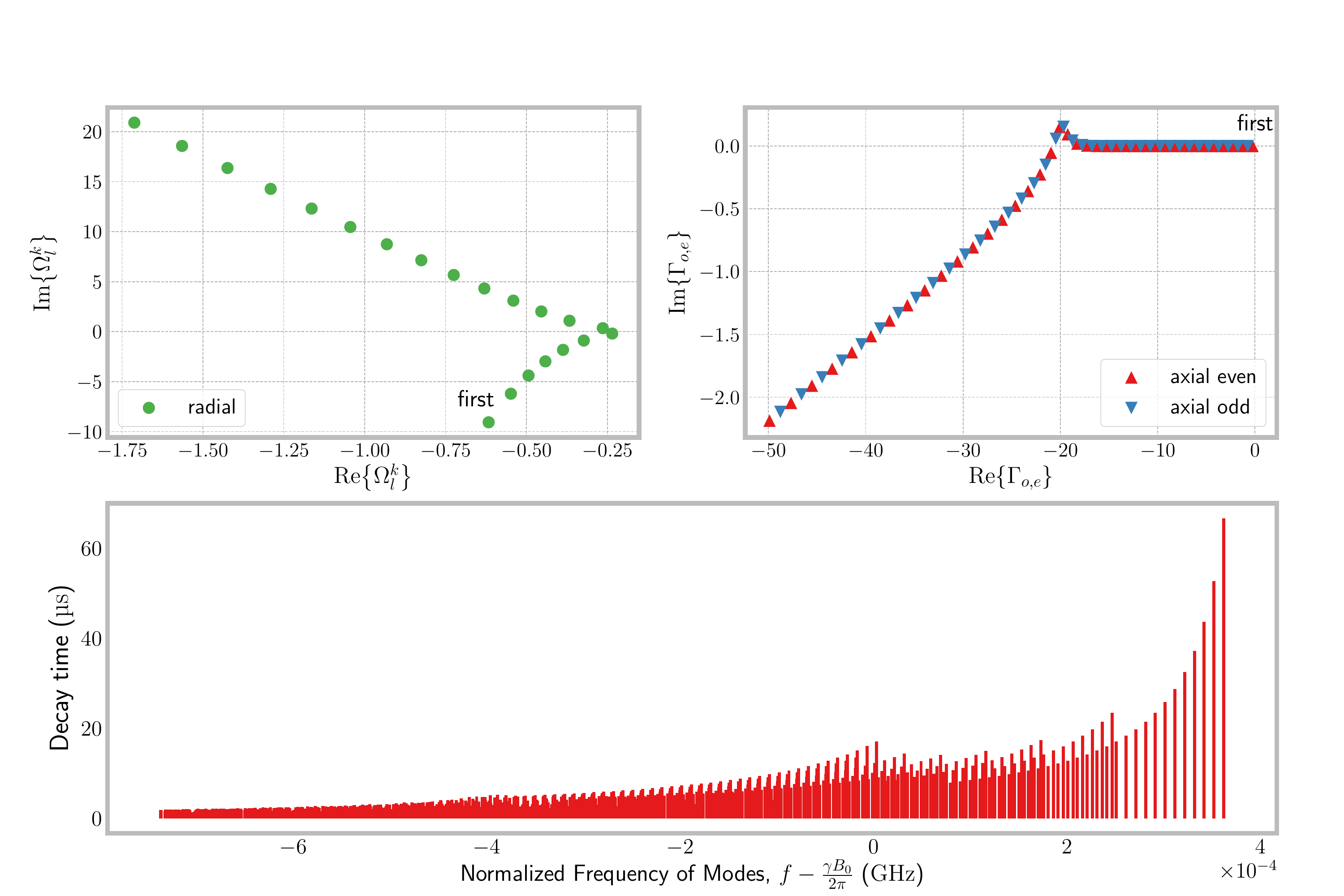}
\caption{Solutions of characteristic equations (upper plots) for the radial (left) and axial (right) mode numbers, and the ISR magnon mode  frequencies (lower plot) for the H gas density $n_H=\SI{2e17}{\per \cubic \centi \meter}$, and temperature $T=\SI{0.25}{\kelvin}$}
\label{fig:roots-and-freqs}
\end{figure*}

\subsection{Characteristics of Magnons}
\begin{figure*}
\includegraphics[width=0.9\textwidth]{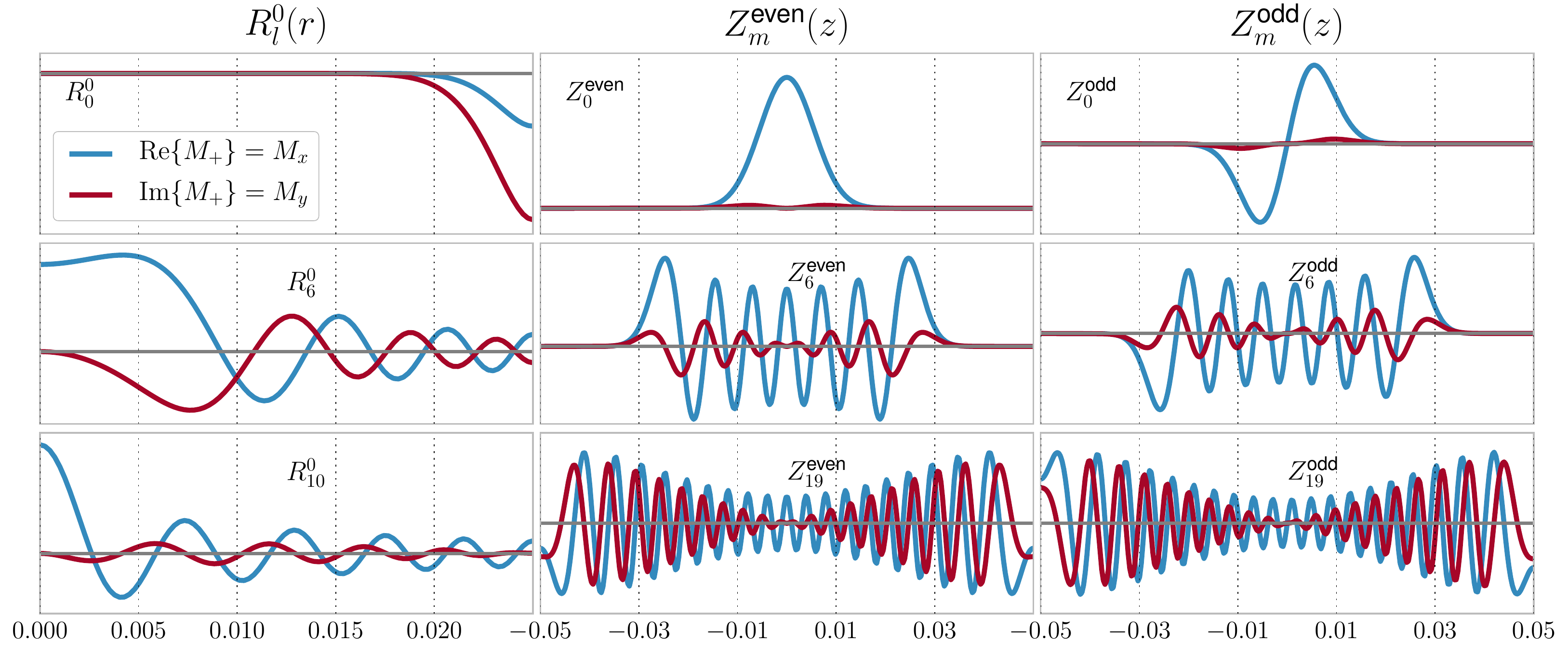}
\caption{Eigenmodes $R^0_{0,6,10}, Z^{\text{even, odd}}_{0, 6,19}, Z^{\text{even, odd}}_{0, 6,19}$}
\label{eigenfunctions}
\end{figure*}

As in the case of a particle in a finite potential, the magnon spectrum splits into two branches: the modes trapped in the saddle potential and in the cylindrical box. The split is readily noticeable in Fig. \ref{fig:roots-and-freqs}, where the behaviour of the roots of the characteristic equations changes at the cusp of the curve. The change can also be seen in the mode functions Fig. (\ref{eigenfunctions}), in general as the expansion of the wave functions beyond the saddle trap. One can also see that the radial numbers have the strongest influence on the mode frequency. However, the difference in frequency between nearest modes even in this case does not exceed $\Delta f \sim 10 kHz$ which is equivalent to $\approx 3$ mG. Therefore, we do not expect to resolve individual modes in the ESR spectrum but rather see an envelope of a large number of modes.  

The fact that the magnons of hydrogen are high-field seekers is evident from Fig. \ref{fig:roots-and-freqs}, where the highest frequency corresponds to the highest magnetic field via $\omega = \gamma B$. The mode functions ($R_0^0, Z_0$ in Fig \ref{eigenfunctions}) are also concentrated in areas of strong magnetic field. 

Evaluating the imaginary part of the frequencies $\omega$ for each mode gives their damping rates, and characteristic life-time of the modes. The magnon decay times are presented in Fig. \ref{fig:roots-and-freqs}). These evaluations are in reasonable agreement with the long lasting oscillations (Fig. \ref{fig:FID vs density}) observed in high density pulsed ESR spectra from the trapped magnons region.

One would expect decay times to decrease as a function of mode number, and generally this seems to be case. However, there are some curious exceptions to the rule: The $(0, l=6, 0, \text{even})$  mode, with frequency close to the saddle point at $\frac{\gamma B_0}{2 \pi}$, has a lifetime longer than many higher frequency modes ($\tau=\SI{17.11}{\micro \second}$, being the 17th longest-lived mode, while  $(0, l=3, 0, \text{even})$ is 27th). This mode seems to be concentrated in the center of the cylinder in  both radial and axial directions (see Fig. \ref{eigenfunctions}).

\section{Discussion}
In the above presented results we observed several features which cannot be explained via classical theory of electron spin resonance signals and lineshapes. First we summarize these observations.

We observed an extra peak in the ESR spectrum originating from the local maximum of magnetic field and located in the region of vanishingly small rf excitation field. We attributed this peak to the electron spin wave modes trapped in the minimum of magnetic potential. We found that a sharp and strong feature emerges from this trapped magnon peak at densities exceeding some critical value. The phase difference of $\approx \pi/2$ is observed for this peak from that of the main ESR signal, which is also consistent with its spatial location at the maximum of the magnetic field. The width of the peak is substantially smaller than the ESR linewidth expected for such an inhomogeneity of static magnetic field. The value of critical density depends on the geometry and shape of the magnetic potential. The saddle potential provides less tight confinement for the spin wave modes than the strong linear gradients combined with the physical walls. The latter case is realized in the bubble experiments, where we found that the critical density is nearly order of magnitude larger (see Fig. \ref{fig:Critical densities}). Analysing the shape of the free induction decays (in the time-domain) we found that above the critical density, they contain two oscillating components. First is the rapidly decaying signal seen also at low densities caused by the magnetic field inhomogeneities. In addition above the critical density a second component appears, which grows after the excitation pulse, reaches maximum, and then slowly decays resulting in a narrow peak in the Fourier transform of the free induction decay. In the first case the coherence is lost because of the differencies in the oscillation frequencies of the large number of magnon modes from which it is composed. The second signal exhibits features of spontaneous coherence developed by the reorganization of the spins within the sample instead of the induced coherence by the excitation pulse. This observation is very similar to the oscillations of the Homogeneously Precessing Domain in liquid $^3$He, recently interpreted in terms of the BEC of magnons \cite{Borovik-Romanov_1984, Bunkov_2010, Autti_2012}.

Using pulsed ESR spectroscopy with the narrow selective excitation we demonstrated that the spin perturbation created in the maximum of the rf field is rapidly transferred into the saddle potential and results in a narrow and strong peak corresponding to the single spin wave mode oscillation from this region. We determined that the spin transport from the excitation region to the trapped state occurs with a speed much faster than ordinary physical diffusion.

We suggest an interpretation of the above mentioned results in terms of a spontaineous coherent oscillation of the lowest energy mode of magnons in the trapping potential. This phenomenon is similar to the giant occupation of the ground state and coherence which occur in Bose-Einstein condensation of atoms obeying Bose statistics. The magnons are bosons, but they are quasi-particles or quanta of excitations without physical mass. The possibility of BEC-like behavior for massless quasiparticles has been predicted by H. Fr{\"o}hlich \cite{Frohlich_1968}, and later developed by A. Bugrij and V. Loktev \citep{Bugrij_2008} and V. Safonov \cite{Safonov_V_2013}. In thermal equilibrium occupation numbers of quasiparticles of energy levels $\epsilon_k$ are given by the Bose distribution function:
\begin{equation}
n_k =  \frac{1}{e^{\frac{\epsilon_k -\mu}{k_B T}}-1} 
\label{eq:Bose distribution}
\end{equation}
with zero chemical potential  $\mu =0$, the well-known Plank distribution. Lowering the temperature will not lead to any kind of condensation; the occupation numbers will just decrease, and no quasiparticles will be found at $T=0$. If however, we turn on an external pumping source which will inject extra quasiparticles the chemical potential will deviate from zero, and will grow as a monotonic function of the pumping rate $I_p$ \cite{Frohlich_1968, Bugrij_2008}. By eventually injecting more and more quasiparticles, we can reach the situation where the chemical potential gets very close to the ground state energy. As follows from Eq.\ref{eq:Bose distribution}, the occupation of the ground state diverges at $\mu = \epsilon_0$ as
\begin{equation}
n_0 =  \frac{k_B T}{\epsilon_0 -\mu} 
\label{eq:n0}
\end{equation}
In order to determine the critical pumping rate, we need to find out how the chemical potential depends on the pumping rate. In the high temperature limit $k_B T\gg \epsilon_0$, which is well justified in our case, the chemical potential is given by the equation \cite{Bugrij_2008}:
\begin{equation}
\mu = I_p \tau_1 \frac{\epsilon_{0}^{2}}{k_B T} 
\label{eq:mu}
\end{equation}
where the dissipation rate of the pumped magnons into the thermal bath  $\tau_{1}^{-1}$ is slow enough compared with the inter-state relaxation rate between quasiparticles in different quantum states $\tau_{2}^{-1}$. The latter condition is required to ensure that the Bose distribution can be applied for the system with pumped magnons. One can see from Eq. \ref{eq:mu} that in order to get $\mu=\epsilon_0$ one needs to provide a pumping rate strong enough, such that $I_p \tau_1= \frac{k_B T}{\epsilon_0} \gg 1$, since we assumed a high temperature approximation. Therefore, the condition for BEC of quasiparticles can be attained if the pumping rate is much faster than relaxation into the thermal bath. For magnons in ferromagnetic films this has been realized  even at room temperature \cite{Demokritov_2006}.

In the case of ISR magnons in H gas the situation gets somewhat more complicated because the effective mass of e-magnons depends on the H gas density as it follows from Eq.\ref{eq:isre}, taking into account that $D_0\sim 1/n_{H}$    \cite{Lhuillier_1982_I, Johnson_1984, Levy_1984}. Therefore, the energies of magnon states  also depend on the H gas density. For the case of harmonic trap this would be $\epsilon_0 \sim (m^{*})^{-1/2} \sim n_{H}^{-1/2}$ . For the real trapping potentials the saddle and cylindrical geometry used in our work we calculated $\epsilon_0$ as a function of $n_H$  numerically using Eq. \ref{eq:isre}. Results are presented in Fig. \ref{fig:Critical densities}. It turns out that $\epsilon_0$  follows $\sim n_{H}^{-1/2}$ dependence quite well. Next, we should take into account that the pumping rate of magnons is proportional to the number of flipped spins in the ESR absorption, which is proportional to the gas density $n_H$. Then, for fixed RF excitation power and trapping geometry we obtain from Eq. \ref{eq:mu} that the chemical potential does not depend on $n_H$. Note, that the pumping rate and the chemical potential under pumping is different for each trapping potential. The ground state energy $\epsilon_0$  for each trap decreases when the H gas density decreases, and at some stage becomes equal to the chemical potential $\mu$. 

In the experiments, we keep the RF power fixed, but follow the ESR signal changes for decreasing (cylindrical geometry) or increasing (bubble) H gas density.  Taking the values of the critical H gas density from Fig. \ref{fig:Critical densities} we find values of the ground state frequency for which the condensation occurs. This happens when the ground state energy becomes equal to the chemical potential $\mu = \epsilon_0$. Condensation occurs earlier for the saddle potential because its ground state energy is lower than that for the bubbles. The fact that the condensation occurs at nearly the same value of $\epsilon_0$ indicates that the magnon pumping rates are nearly same for both traps even though the H gas densities differ by an order of magnitude. The RF field intensity for the saddle trap is substantially smaller than that for the bubble, which is compensated for the much larger volume and the density of states in the saddle trap, finally providing nearly the same pumping rate.

\section{Conclusions and future prospects}
	In this work we demonstrated that magnons in a quantum gas of atomic hydrogen can be trapped in the potential well created by local maximum of magnetic field and the walls of an experimental cell. A large variety of spin wave modes were observed as modulations of the ESR spectra. At a high density of H gas, a strong and narrow peak in the ESR spectrum emerged, which is caused by the strong and coherent oscillations of the ground state mode in the trap. This phenomenon can be explained in terms of the Bose-Einstein condensation of magnons. A very interesting topic for continuing experiments would be to study effects related to interactions between magnons and possible spin superfluidity. In the future we plan to improve control of the magnetic field in the trap and utilize two coupled traps for studying interference and Josephson tunnelling effects between two magnon condensates.

\begin{acknowledgments}
This work was supported by the Academy of Finland (Grants No. 122595, and 133682), the Wihuri Foundation and US NSF grant DMR 1707565. 
\end{acknowledgments}

\appendix
\section{Experimental details}

\subsection*{Construction of the sample cell}
\begin{figure}

\includegraphics[width=8 cm]{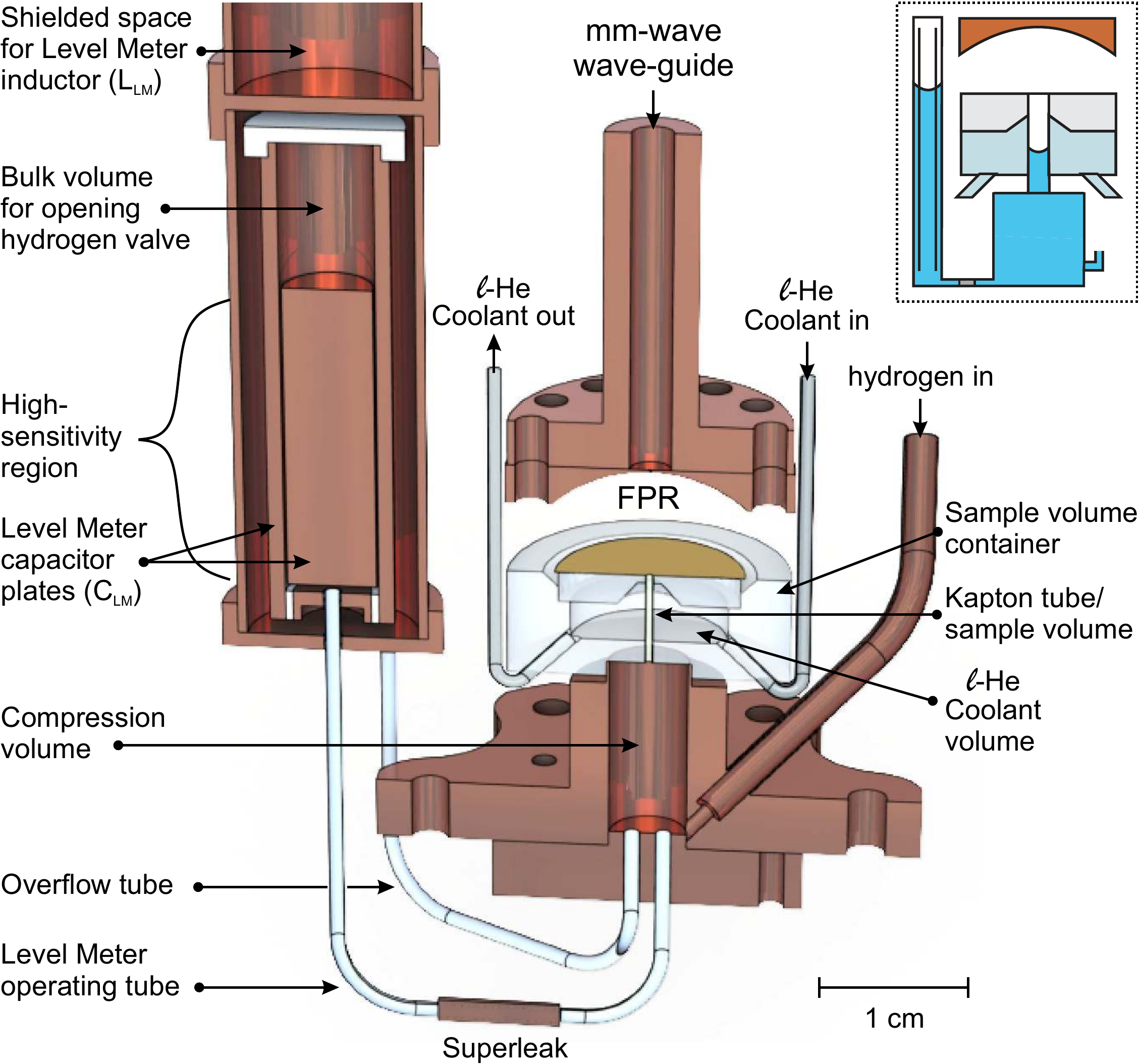}
\caption{Cross section of the experimental cell. Helium piston reservoir with level meter capacitor, FP resonator for ESR, coolant helium volume and tubes, hydrogen gas sample accumulation volume and fill line, compression volume and excitation region.}\label{fig:exp_cell_cross_section}
\end{figure}

Presence of the physical walls in experiments with atomic hydrogen is the main obstacle for reaching high densities of the gas because of the adsorption and subsequent recombination of atoms on the surface. Superfluid helium turned out to be the material with smallest adsorption energy of $\sim$1 K, and therefore, all experiments with high density H gas have to be done in the sample cell covered by superfluid helium and at temperatures above $\sim$300 mK. The presence of superfluid helium allows realization of a simple compression technique with a helium piston driven by the fountain pressure of superfluid (see Figs. \ref{fig:cell_1}, \ref{fig:exp_cell_cross_section}).

The position of the helium piston and the compression process is controlled by the heater in the left column of the U-tube system, with the two columns being separated by a superleak. Raising the temperature of the left column leads to an increase of its height caused by the temperature dependent fountain pressure of the superfluid helium. Helium is pumped back into the compression/sample volume by ramping down the heat applied to the left column.  With such a fountain pump technique we compress H gas to the top of the 0.5 mm diameter cylinder, reducing its volume by nearly three orders of magnitude to reach densities exceeding 10$^{18}$ cm$^{-3}$. The heat ramp rate is carefully adjusted at each stage of the compression cycle in order to get eventually a maximum density and well defined geometry. Too fast compression may lead to an explosive recombination of the sample. To ensure effective cooling, the plastic tube with compressed gas is immersed into a chamber with continuous flow of superfluid helium which passes through a special heat exchanger at the mixing chamber of a dilution refrigerator. Using a capacitance gauge in the left column we can measure the position of the helium meniscus in the right column with sub-micrometer resolution.  At the end of the compression cycle, the temperatures of the columns are stabilized, and the height of the cylinder with compressed H gas is defined by the balance between the fountain pressure plus the hydraulic head pressure of helium, and the hydrogen gas pressure. 

The sample volume is separated from the Fabry-Perot cavity and vacuum space of the dilution refigerator with a gold coated kapton film, which prevents the hydrogen from leaking into the high amplitude ESR-volume inside the cavity. The gold coating acts as the planar mirror of the cavity and allows the rf-field to interact with the sample through a sub-critical (0.4 mm) hole in the gold coating, centred with the sample volume. This construction produces a highly inhomogeneous rf-field in the sample volume, with a characteristic $(1/e)$ penetration length of $\approx 80$ $\mu$m into the sample volume. An inhomogeneous ESR-excitation is crucial for the excitation of spin wave modes.

\begin{figure}
\includegraphics[width=8 cm]{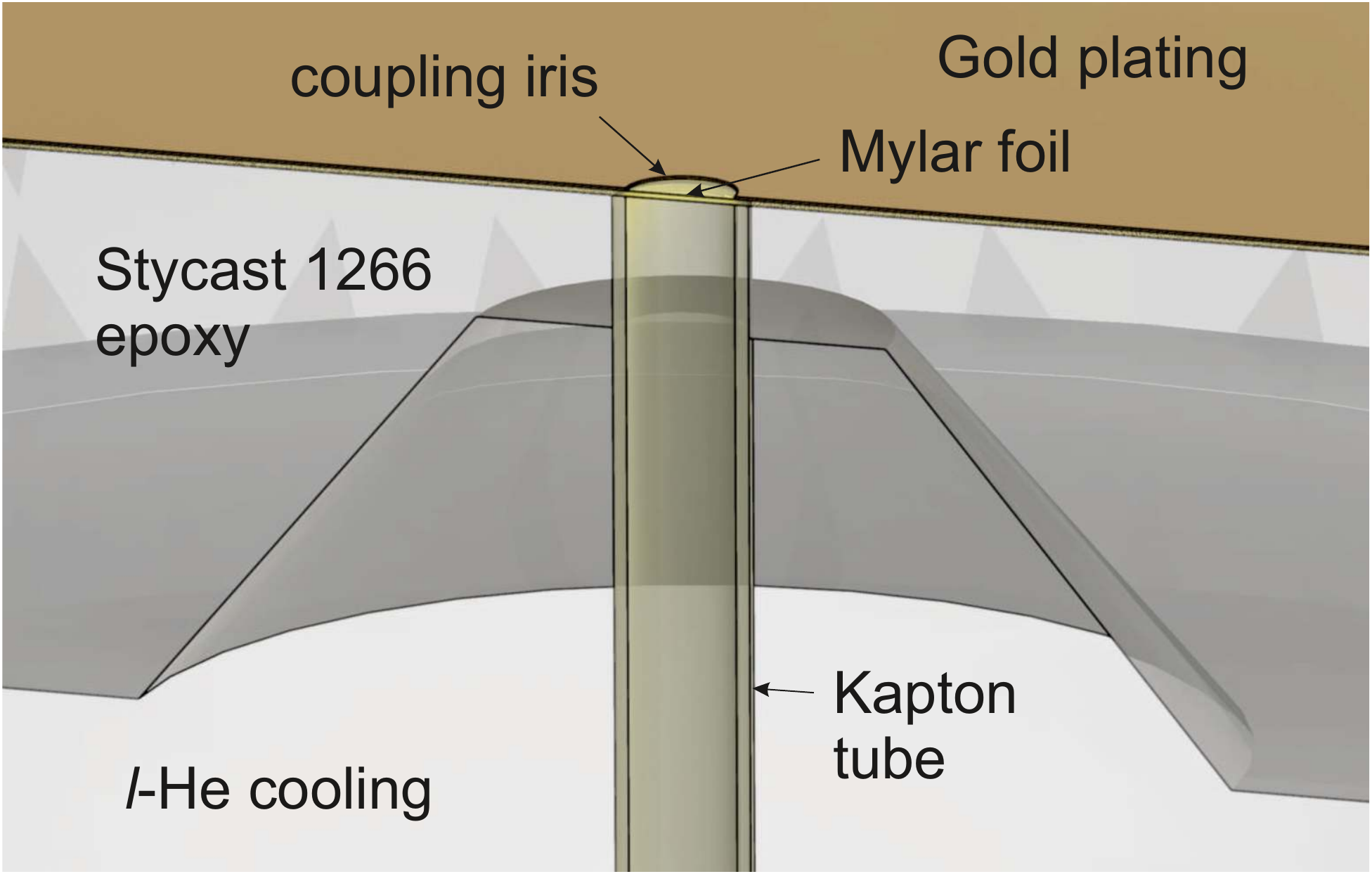}
\caption{Magnified cross section of the experimental cell around the top part of the sample volume (cylidrical volume in the centre). Also visible are the stycast meniscus, gold coated mirror of the Fabry-Perot resonator and the coupling hole in the gold coating above the sample volume, and the top part of the helium coolant volume.}\label{fig:exp_cell_cross_section_magnification}
\end{figure}

\subsection*{Experimental procedure}

A liquid helium piston was used to compress the hydrogen gas in order to reach high densities. The hydrogen gas was first loaded into a larger volume by using the liquid helium piston as a valve opening and closing the hydrogen gas loading inlet.

The liquid helium piston was operated by adjusting a temperature difference ($\Delta$T) between two sides of a super leak. On the reservoir side of the piston the liquid helium was stored between coaxially aligned capacitor plates. Above the inner capacitor plate was a larger bulk volume, which allowed emptying the compression volume for loading a hydrogen gas sample. The capacitor was part of an LC-circuit driven by a tunnel diode circuitry. The operating frequency of $\approx 25$ MHz of this LC-oscillator was detected at room temperature. The frequency difference between the empty and full (of liquid helium) capacitor was $\approx 190$ kHz and with our usual sampling rate of 0.5 Hz the detection accuracy was a few Hertz. This gave us a volume resolution of approximately 1 nl corresponding to a helium level change of $\approx \SI{0.2}{\micro \meter}$ in the sample volume.
\begin{figure}
\includegraphics[width=8 cm]{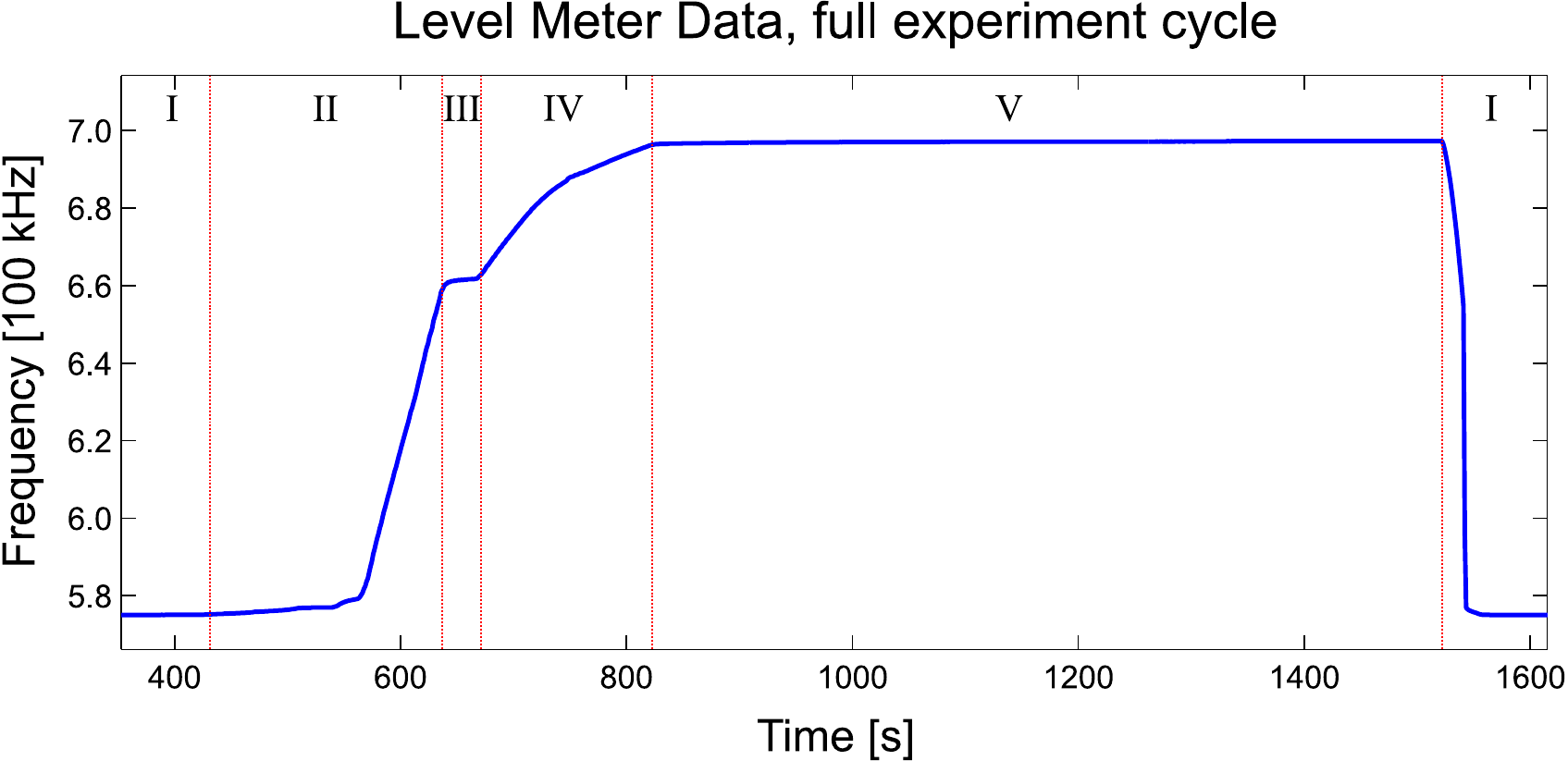}
\caption{Level meter data for a full experiment cycle}\label{fig:LM_data_full_cycle}
\end{figure}

\begin{figure}
\includegraphics[width=8 cm]{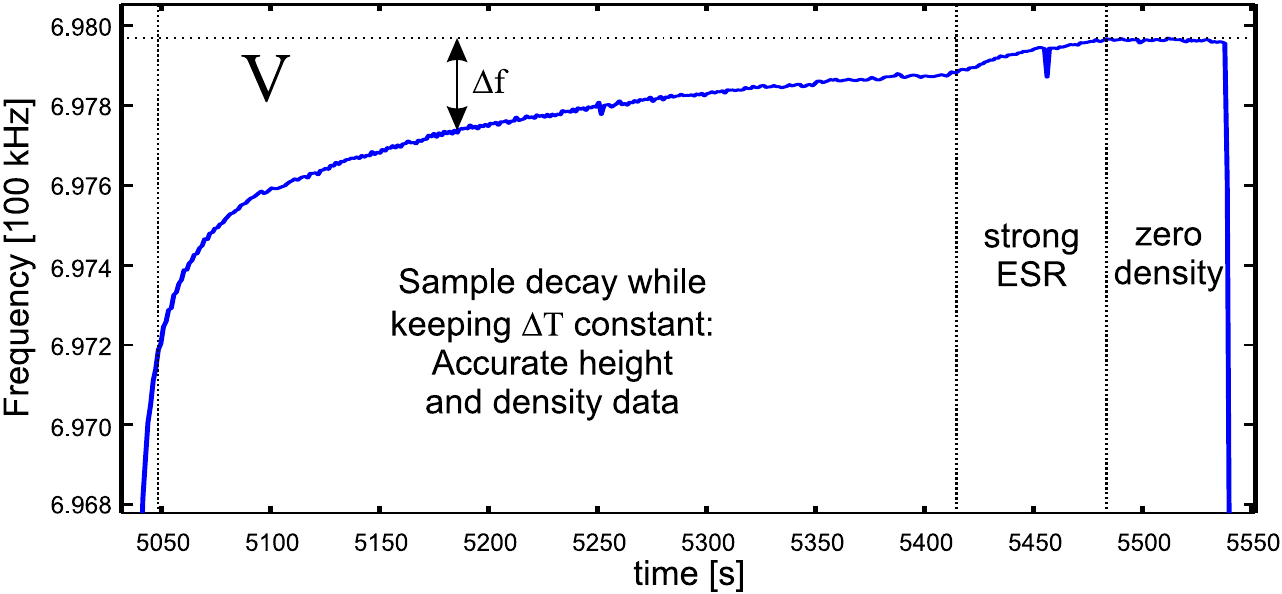}
\caption{Level meter data detail for sample decay}\label{fig:LM_data_sample_decay}
\end{figure}

A typical experiment cycle lasted $\approx 1500$ seconds and consisted of five distinctive stages: I Accumulation, II Pre-compression, III Pause, IV Final compression and V Sample Decay.

The recording of spectra already started during the accumulation stage I with low hydrogen gas densities, while ISRE-induced spin dynamics were still absent. The line shape and width were defined by the finite interaction time of the atoms with the rf excitation field during their ballistic flight through the EF region. The maximum half-width of $\approx 50 $ mG ($\sim 150$ kHz) well matches the $\approx 6 \mu$s flight time across the $\approx 0.4$ mm diameter EF. The width of the ESR lines rapidly decreased with increase of H gas density, clearly indicating on the transition from the molecular to diffusive region of motion. Minimum width of $\approx 20$ mG was reached at densities exceeding $\sim 10^{17}$ cm$^{-3}$ as defined by the residual inhomogeneity of static magnetic field in the EF.  The narrow low density spectra served as calibration against long term frequency/magnetic field drifts.

Initially accumulated sample gas included both high-field seeking hyperfine states \textit{a} ($\vert m_I, m_S \rangle = \vert +1/2, -1/2 \rangle$) and \textit{b} ($\vert m_I, m_S \rangle = \vert -1/2, -1/2 \rangle$). Increase of density at the pre-compression stage leads to a preferential recombination of the \textit{a}-state (see e.g. \cite{BlueBible} for details) resulting to a substantial extra heat. A pause in the compression sequence was found to be necessary, in order to accommodate this process. A too fast compression with a-state atoms in the sample lead to a runaway recombination and explosion of the sample. After few tens of seconds of steady state evolution, the final compression stage was initiated, raising the temperature difference between compression and reservoir volumes to a desired value. The range of possible final compression forces was large leading to a large range final hydrogen gas densities, although too strong compressions again induced a runaway recombination, irrespective of the pause before the final compression.
\begin{figure*}
\includegraphics[width=1\textwidth]{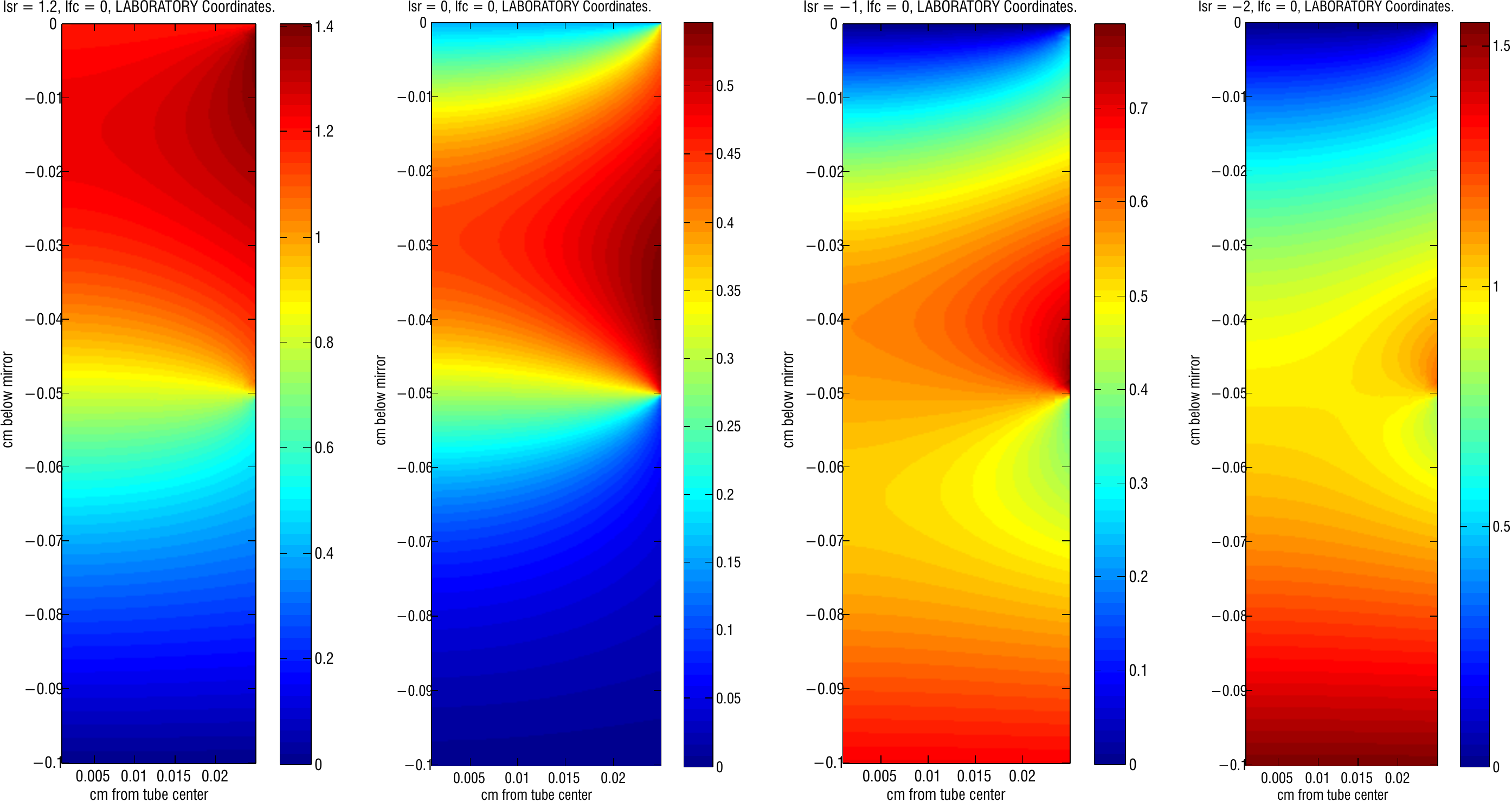}
\caption{Half cross-sections of magnetic field amplitudes near the top of the sample volume for four different gradient coil current values. The added gradient field is homogeneous in radial direction, adding only a variable, linear gradient in z-dircetion. The local magnetic field maximum (dark red) at the outer wall of the sample volume moves both spatially and in relative field strength compared to the excitation and detection region at the top of the volume. }\label{fig:Magnetid_field_gradients}
\end{figure*}

During the compression the geometry of the sample volume is constantly changing, as the upper meniscus of the liquid helium piston rises in the volume, pushing the hydrogen gas towards the top of the volume. As most of the detectable spin wave dynamics is happening at the top most mm of the sample volume, the helium piston induced geometry changes affect the spectra only at the very end of a sufficiently strong compression. During weak compressions the piston does not reach the top of the sample volume. For the weak compression, the highest hydrogen gas densities occur at the end of the compression volume heating ramp, after which the hydrogen gas density starts to decrease due to recombination and ending with a vacuum above the piston. The most accurate hydrogen gas density measurements were performed during the period of constant $\Delta$T between the volumes, when the hydrogen gas density was declining. The density dependence of the spin wave dynamics in a cylindrical geometry, and with a saddle potential minimum for the magnons, was seen twice during these compressions. Once while the density is increasing during the final compression, and in reversed order during the decay of the sample.

At the end of relatively strong final compressions the hydrogen gas collapsed into a bubble immersed in liquid helium at the top of the sample volume. The bubble stage of the compression usually lasted a few tens of seconds, during which we could measure tens of spectra. Although we were capable of measuring the volume changes relatively accurately even during the final seconds of the bubble life-time, determining the hydrogen pressure turned out be non-trivial. In contrast to the weaker compression, the hydrogen density is monotonically increasing during the evolution of the bubble.

\subsection*{Magnetic Field Gradients}
The total magnetic field inside the sample volume is a sum of fields from several independent sources. The overall direction of the field is set by the main coil of the superconducting magnet, creating the 4,6 Tesla polarizing field. In our experiments this polarizing field is parallel to the symmetry axis of the experimental cell. A strong magnetic field is needed to maintain the spin-polarization of the hydrogen atoms. Spin polarization is necessary as it significantly slows down the recombination into molecular hydrogen, allowing experiments with atomic hydrogen.

Magnetic fields from all other sources, including the built in shim, gradient, and sweep coils of the main magnet, are much weaker than the main coil field, but have a significant role in fine tuning the total field amplitude and modulating the amplitude gradients. In addition to the built-in coils in the main magnet, we had altogether 5 external coils, of which two were in a Helmholtz configuration, and other two in an anti-Helmholtz configuration. Due to the large difference in field amplitudes between the main coil and the field from the rest of the coils, only the field component parallel to the main coil field affects the total field amplitude. The transverse components introduce only a vanishingly small tilt to the direction of the total magnetic field.

Besides the magnetic field from the externally controlled coils, magnetized material in the vicinity of the sample volume affects the total magnetic field amplitude. In these experiments the slight diamagnetic magnetization of the epoxy (Stycast 1266) collar around the top of the sample volume created a toroidal magnetic field maximum at the outer wall of the sample volume. The location and relative strength of this maximum, compared to the magnetic field at the excitation region at the top center of the sample volume, was controllable by varying the vertical field gradient via coils. In fig. \ref{fig:Magnetid_field_gradients} the simulated magnetic field amplitude for four different gradient coil currents is shown. 

\section{Characteristic Equations} The characteristic equations shown below impose the boundary conditions on the modes and allow solving for $\Omega_l^k, \Gamma_m^{\text{even, odd}}$. For hydrogen these are
\begin{widetext}
\begin{multline*}
R^{2}\sqrt{-\mu+i}\left(2\Omega+k+1\right)\pFq{1}{1}{\Omega+\frac{k+3}{2}}{k+2}{\frac{R^{2}}{\lambda_{r}^{2}}\sqrt{-\mu+i}}\\+\left(k+1\right)\left(k\lambda_{r}^{2}-R^{2}\sqrt{-\mu+i}\right)\pFq{1}{1}{\Omega+\frac{k+1}{2}}{k+1}{\frac{R^{2}}{\lambda_{r}^{2}}\sqrt{-\mu+i}}=0
\end{multline*}
\begin{gather*}
\left(4\Gamma_{e}+1\right)\pFq{1}{1}{\Gamma_{e}+\frac{5}{4}}{\frac{3}{2}}{\frac{L^{2}}{\lambda_{z}^{2}}\sqrt{-\left(-\mu+i\right)}}-\pFq{1}{1}{\Gamma_{e}+\frac{1}{4}}{\frac{1}{2}}{\frac{L^{2}}{\lambda_{z}^{2}}\sqrt{-\left(-\mu+i\right)}}=0
 \\
\frac{L^{2}}{3}\left(4\Gamma_{o}+3\right)\sqrt{-\left(-\mu+i\right)}\pFq{1}{1}{\Gamma_{o}+\frac{7}{4}}{\frac{5}{2}}{\frac{L^{2}}{\lambda_{z}^{2}}\sqrt{-\left(-\mu+i\right)}}+\left(\lambda_{z}^{2}-L^{2}\sqrt{-\left(-\mu+i\right)}\right)\pFq{1}{1}{\Gamma_{o}+\frac{3}{4}}{\frac{4}{2}}{\frac{L^{2}}{\lambda_{z}^{2}}\sqrt{-\left(-\mu+i\right)}}=0
\end{gather*}
\end{widetext}

\bibliographystyle{apsrev4-1}
\bibliography{MagLong}

%merlin.mbs apsrev4-1.bst 2010-07-25 4.21a (PWD, AO, DPC) hacked
%Control: key (0)
%Control: author (72) initials jnrlst
%Control: editor formatted (1) identically to author
%Control: production of article title (-1) disabled
%Control: page (0) single
%Control: year (1) truncated
%Control: production of eprint (0) enabled
\begin{thebibliography}{28}%
\makeatletter
\providecommand \@ifxundefined [1]{%
 \@ifx{#1\undefined}
}%
\providecommand \@ifnum [1]{%
 \ifnum #1\expandafter \@firstoftwo
 \else \expandafter \@secondoftwo
 \fi
}%
\providecommand \@ifx [1]{%
 \ifx #1\expandafter \@firstoftwo
 \else \expandafter \@secondoftwo
 \fi
}%
\providecommand \natexlab [1]{#1}%
\providecommand \enquote  [1]{``#1''}%
\providecommand \bibnamefont  [1]{#1}%
\providecommand \bibfnamefont [1]{#1}%
\providecommand \citenamefont [1]{#1}%
\providecommand \href@noop [0]{\@secondoftwo}%
\providecommand \href [0]{\begingroup \@sanitize@url \@href}%
\providecommand \@href[1]{\@@startlink{#1}\@@href}%
\providecommand \@@href[1]{\endgroup#1\@@endlink}%
\providecommand \@sanitize@url [0]{\catcode `\\12\catcode `\$12\catcode
  `\&12\catcode `\#12\catcode `\^12\catcode `\_12\catcode `\%12\relax}%
\providecommand \@@startlink[1]{}%
\providecommand \@@endlink[0]{}%
\providecommand \url  [0]{\begingroup\@sanitize@url \@url }%
\providecommand \@url [1]{\endgroup\@href {#1}{\urlprefix }}%
\providecommand \urlprefix  [0]{URL }%
\providecommand \Eprint [0]{\href }%
\providecommand \doibase [0]{http://dx.doi.org/}%
\providecommand \selectlanguage [0]{\@gobble}%
\providecommand \bibinfo  [0]{\@secondoftwo}%
\providecommand \bibfield  [0]{\@secondoftwo}%
\providecommand \translation [1]{[#1]}%
\providecommand \BibitemOpen [0]{}%
\providecommand \bibitemStop [0]{}%
\providecommand \bibitemNoStop [0]{.\EOS\space}%
\providecommand \EOS [0]{\spacefactor3000\relax}%
\providecommand \BibitemShut  [1]{\csname bibitem#1\endcsname}%
\let\auto@bib@innerbib\@empty
%</preamble>
\bibitem [{\citenamefont {Bose}(1924)}]{Bose_1924}%
  \BibitemOpen
  \bibfield  {author} {\bibinfo {author} {\bibfnamefont {S.~N.}\ \bibnamefont
  {Bose}},\ }\href@noop {} {\bibfield  {journal} {\bibinfo  {journal}
  {Zeitschrift f{\"{u}}r Physik A Hadrons and Nuclei}\ }\textbf {\bibinfo
  {volume} {26}},\ \bibinfo {pages} {178} (\bibinfo {year} {1924})}\BibitemShut
  {NoStop}%
\bibitem [{\citenamefont {Einstein}(1924)}]{Einstein_1924}%
  \BibitemOpen
  \bibfield  {author} {\bibinfo {author} {\bibfnamefont {A.}~\bibnamefont
  {Einstein}},\ }\href@noop {} {\bibfield  {journal} {\bibinfo  {journal}
  {Sitzungsber. Kgl. Preuss. Akad. Wiss.}\ }\textbf {\bibinfo {volume} {261}}
  (\bibinfo {year} {1924})}\BibitemShut {NoStop}%
\bibitem [{\citenamefont {Anderson}\ \emph {et~al.}(1995)\citenamefont
  {Anderson}, \citenamefont {Ensher}, \citenamefont {Matthews}, \citenamefont
  {Wieman},\ and\ \citenamefont {Cornell}}]{Anderson_1995}%
  \BibitemOpen
  \bibfield  {author} {\bibinfo {author} {\bibfnamefont {M.~H.}\ \bibnamefont
  {Anderson}}, \bibinfo {author} {\bibfnamefont {J.~R.}\ \bibnamefont
  {Ensher}}, \bibinfo {author} {\bibfnamefont {M.~R.}\ \bibnamefont
  {Matthews}}, \bibinfo {author} {\bibfnamefont {C.~E.}\ \bibnamefont
  {Wieman}}, \ and\ \bibinfo {author} {\bibfnamefont {E.~A.}\ \bibnamefont
  {Cornell}},\ }\href@noop {} {\bibfield  {journal} {\bibinfo  {journal}
  {Science}\ }\textbf {\bibinfo {volume} {269}},\ \bibinfo {pages} {198}
  (\bibinfo {year} {1995})}\BibitemShut {NoStop}%
\bibitem [{\citenamefont {Davis}\ \emph {et~al.}(1995)\citenamefont {Davis},
  \citenamefont {Mewes}, \citenamefont {Andrews}, \citenamefont {van Druten},
  \citenamefont {Durfee}, \citenamefont {Kurn},\ and\ \citenamefont
  {Ketterle}}]{Davis_1995}%
  \BibitemOpen
  \bibfield  {author} {\bibinfo {author} {\bibfnamefont {K.~B.}\ \bibnamefont
  {Davis}}, \bibinfo {author} {\bibfnamefont {M.-O.}\ \bibnamefont {Mewes}},
  \bibinfo {author} {\bibfnamefont {M.~R.}\ \bibnamefont {Andrews}}, \bibinfo
  {author} {\bibfnamefont {N.~J.}\ \bibnamefont {van Druten}}, \bibinfo
  {author} {\bibfnamefont {D.~S.}\ \bibnamefont {Durfee}}, \bibinfo {author}
  {\bibfnamefont {D.~M.}\ \bibnamefont {Kurn}}, \ and\ \bibinfo {author}
  {\bibfnamefont {W.}~\bibnamefont {Ketterle}},\ }\href@noop {} {\bibfield
  {journal} {\bibinfo  {journal} {Physical Review Letters}\ }\textbf {\bibinfo
  {volume} {75}},\ \bibinfo {pages} {3969} (\bibinfo {year}
  {1995})}\BibitemShut {NoStop}%
\bibitem [{\citenamefont {Frohlich}(1968)}]{Frohlich_1968}%
  \BibitemOpen
  \bibfield  {author} {\bibinfo {author} {\bibfnamefont {H.}~\bibnamefont
  {Frohlich}},\ }\href@noop {} {\bibfield  {journal} {\bibinfo  {journal}
  {Physics Letters}\ }\textbf {\bibinfo {volume} {26A}},\ \bibinfo {pages}
  {402} (\bibinfo {year} {1968})}\BibitemShut {NoStop}%
\bibitem [{\citenamefont {Kasprzak}\ \emph {et~al.}(2006)\citenamefont
  {Kasprzak} \emph {et~al.}}]{Kasprzak_2006}%
  \BibitemOpen
  \bibfield  {author} {\bibinfo {author} {\bibfnamefont {J.}~\bibnamefont
  {Kasprzak}} \emph {et~al.},\ }\href@noop {} {\bibfield  {journal} {\bibinfo
  {journal} {Nature}\ }\textbf {\bibinfo {volume} {443}},\ \bibinfo {pages}
  {409} (\bibinfo {year} {2006})}\BibitemShut {NoStop}%
\bibitem [{\citenamefont {Ruegg}\ \emph {et~al.}(2003)\citenamefont {Ruegg},
  \citenamefont {Cavadini}, \citenamefont {Furrer}, \citenamefont {Güdel},
  \citenamefont {Krämer}, \citenamefont {Mutka}, \citenamefont {Wildes},
  \citenamefont {Habicht},\ and\ \citenamefont {Vorderwisch}}]{Ruegg_2003}%
  \BibitemOpen
  \bibfield  {author} {\bibinfo {author} {\bibfnamefont {C.}~\bibnamefont
  {Ruegg}}, \bibinfo {author} {\bibfnamefont {N.}~\bibnamefont {Cavadini}},
  \bibinfo {author} {\bibfnamefont {A.}~\bibnamefont {Furrer}}, \bibinfo
  {author} {\bibfnamefont {H.~U.}\ \bibnamefont {Güdel}}, \bibinfo {author}
  {\bibfnamefont {K.}~\bibnamefont {Krämer}}, \bibinfo {author} {\bibfnamefont
  {H.}~\bibnamefont {Mutka}}, \bibinfo {author} {\bibfnamefont
  {A.}~\bibnamefont {Wildes}}, \bibinfo {author} {\bibfnamefont
  {K.}~\bibnamefont {Habicht}}, \ and\ \bibinfo {author} {\bibfnamefont
  {P.}~\bibnamefont {Vorderwisch}},\ }\href@noop {} {\bibfield  {journal}
  {\bibinfo  {journal} {Nature}\ }\textbf {\bibinfo {volume} {423}},\ \bibinfo
  {pages} {52} (\bibinfo {year} {2003})}\BibitemShut {NoStop}%
\bibitem [{\citenamefont {Demokritov}\ \emph {et~al.}(2006)\citenamefont
  {Demokritov} \emph {et~al.}}]{Demokritov_2006}%
  \BibitemOpen
  \bibfield  {author} {\bibinfo {author} {\bibfnamefont {S.~O.}\ \bibnamefont
  {Demokritov}} \emph {et~al.},\ }\href@noop {} {\bibfield  {journal} {\bibinfo
   {journal} {Nature}\ }\textbf {\bibinfo {volume} {443}},\ \bibinfo {pages}
  {430} (\bibinfo {year} {2006})}\BibitemShut {NoStop}%
\bibitem [{\citenamefont {Autti}\ \emph {et~al.}(2012)\citenamefont {Autti}
  \emph {et~al.}}]{Autti_2012}%
  \BibitemOpen
  \bibfield  {author} {\bibinfo {author} {\bibfnamefont {S.}~\bibnamefont
  {Autti}} \emph {et~al.},\ }\href@noop {} {\bibfield  {journal} {\bibinfo
  {journal} {Physical Review Letters}\ }\textbf {\bibinfo {volume} {108}},\
  \bibinfo {pages} {145303} (\bibinfo {year} {2012})}\BibitemShut {NoStop}%
\bibitem [{\citenamefont {Klaers}\ \emph {et~al.}(2010)\citenamefont {Klaers},
  \citenamefont {Schmitt}, \citenamefont {Vewinger},\ and\ \citenamefont
  {Weitz}}]{Klaers_2010}%
  \BibitemOpen
  \bibfield  {author} {\bibinfo {author} {\bibfnamefont {J.}~\bibnamefont
  {Klaers}}, \bibinfo {author} {\bibfnamefont {J.}~\bibnamefont {Schmitt}},
  \bibinfo {author} {\bibfnamefont {F.}~\bibnamefont {Vewinger}}, \ and\
  \bibinfo {author} {\bibfnamefont {M.}~\bibnamefont {Weitz}},\ }\href@noop {}
  {\bibfield  {journal} {\bibinfo  {journal} {Nature}\ }\textbf {\bibinfo
  {volume} {468}},\ \bibinfo {pages} {545} (\bibinfo {year}
  {2010})}\BibitemShut {NoStop}%
\bibitem [{\citenamefont {Vainio}\ \emph {et~al.}(2015)\citenamefont {Vainio},
  \citenamefont {Ahokas}, \citenamefont {J{\"{a}}rvinen}, \citenamefont
  {Lehtonen}, \citenamefont {Novotny}, \citenamefont {Sheludiakov},
  \citenamefont {Suominen}, \citenamefont {Vasiliev}, \citenamefont {Zvezdov},
  \citenamefont {Khmelenko},\ and\ \citenamefont {Lee}}]{Vainio2015}%
  \BibitemOpen
  \bibfield  {author} {\bibinfo {author} {\bibfnamefont {O.}~\bibnamefont
  {Vainio}}, \bibinfo {author} {\bibfnamefont {J.}~\bibnamefont {Ahokas}},
  \bibinfo {author} {\bibfnamefont {J.}~\bibnamefont {J{\"{a}}rvinen}},
  \bibinfo {author} {\bibfnamefont {L.}~\bibnamefont {Lehtonen}}, \bibinfo
  {author} {\bibfnamefont {S.}~\bibnamefont {Novotny}}, \bibinfo {author}
  {\bibfnamefont {S.}~\bibnamefont {Sheludiakov}}, \bibinfo {author}
  {\bibfnamefont {K.-a.}\ \bibnamefont {Suominen}}, \bibinfo {author}
  {\bibfnamefont {S.}~\bibnamefont {Vasiliev}}, \bibinfo {author}
  {\bibfnamefont {D.}~\bibnamefont {Zvezdov}}, \bibinfo {author} {\bibfnamefont
  {V.~V.}\ \bibnamefont {Khmelenko}}, \ and\ \bibinfo {author} {\bibfnamefont
  {D.~M.}\ \bibnamefont {Lee}},\ }\href {\doibase
  10.1103/PhysRevLett.114.125304} {\bibfield  {journal} {\bibinfo  {journal}
  {Physical Review Letters}\ }\textbf {\bibinfo {volume} {114}},\ \bibinfo
  {pages} {125304} (\bibinfo {year} {2015})}\BibitemShut {NoStop}%
\bibitem [{\citenamefont {Nacher}\ \emph {et~al.}(1984)\citenamefont {Nacher},
  \citenamefont {Tastevin}, \citenamefont {Leduc}, \citenamefont {Crampton},\
  and\ \citenamefont {Lalo{\"{e}}}}]{Nacher_1984}%
  \BibitemOpen
  \bibfield  {author} {\bibinfo {author} {\bibfnamefont {P.~J.}\ \bibnamefont
  {Nacher}}, \bibinfo {author} {\bibfnamefont {G.}~\bibnamefont {Tastevin}},
  \bibinfo {author} {\bibfnamefont {M.}~\bibnamefont {Leduc}}, \bibinfo
  {author} {\bibfnamefont {S.~B.}\ \bibnamefont {Crampton}}, \ and\ \bibinfo
  {author} {\bibfnamefont {F.}~\bibnamefont {Lalo{\"{e}}}},\ }\href@noop {}
  {\bibfield  {journal} {\bibinfo  {journal} {Journal de Physique Lettres}\
  }\textbf {\bibinfo {volume} {45}},\ \bibinfo {pages} {441} (\bibinfo {year}
  {1984})}\BibitemShut {NoStop}%
\bibitem [{\citenamefont {Johnson}\ \emph {et~al.}(1984)\citenamefont
  {Johnson}, \citenamefont {Denker}, \citenamefont {Bigelow}, \citenamefont
  {L{\'{e}}vy}, \citenamefont {Freed},\ and\ \citenamefont
  {Lee}}]{Johnson_1984}%
  \BibitemOpen
  \bibfield  {author} {\bibinfo {author} {\bibfnamefont {B.~R.}\ \bibnamefont
  {Johnson}}, \bibinfo {author} {\bibfnamefont {J.~S.}\ \bibnamefont {Denker}},
  \bibinfo {author} {\bibfnamefont {N.~P.}\ \bibnamefont {Bigelow}}, \bibinfo
  {author} {\bibfnamefont {L.~P.}\ \bibnamefont {L{\'{e}}vy}}, \bibinfo
  {author} {\bibfnamefont {J.~H.}\ \bibnamefont {Freed}}, \ and\ \bibinfo
  {author} {\bibfnamefont {D.~M.}\ \bibnamefont {Lee}},\ }\href@noop {}
  {\bibfield  {journal} {\bibinfo  {journal} {Physical Review Letters}\
  }\textbf {\bibinfo {volume} {52}},\ \bibinfo {pages} {1508} (\bibinfo {year}
  {1984})}\BibitemShut {NoStop}%
\bibitem [{\citenamefont {McGuirk}\ \emph {et~al.}(2002)\citenamefont
  {McGuirk}, \citenamefont {Lewandowski}, \citenamefont {Harber}, \citenamefont
  {Nikuni}, \citenamefont {Williams},\ and\ \citenamefont
  {Cornell}}]{McGuirk_2002}%
  \BibitemOpen
  \bibfield  {author} {\bibinfo {author} {\bibfnamefont {J.~M.}\ \bibnamefont
  {McGuirk}}, \bibinfo {author} {\bibfnamefont {H.~J.}\ \bibnamefont
  {Lewandowski}}, \bibinfo {author} {\bibfnamefont {D.~M.}\ \bibnamefont
  {Harber}}, \bibinfo {author} {\bibfnamefont {T.}~\bibnamefont {Nikuni}},
  \bibinfo {author} {\bibfnamefont {J.~E.}\ \bibnamefont {Williams}}, \ and\
  \bibinfo {author} {\bibfnamefont {E.~A.}\ \bibnamefont {Cornell}},\
  }\href@noop {} {\bibfield  {journal} {\bibinfo  {journal} {Physical Review
  Letters}\ }\textbf {\bibinfo {volume} {89}},\ \bibinfo {pages} {90402}
  (\bibinfo {year} {2002})}\BibitemShut {NoStop}%
\bibitem [{\citenamefont {Vainio}\ \emph
  {et~al.}(2012{\natexlab{a}})\citenamefont {Vainio}, \citenamefont {Ahokas},
  \citenamefont {Novotny}, \citenamefont {Sheludyakov}, \citenamefont
  {Zvezdov}, \citenamefont {Suominen},\ and\ \citenamefont
  {Vasiliev}}]{Vainio2012}%
  \BibitemOpen
  \bibfield  {author} {\bibinfo {author} {\bibfnamefont {O.}~\bibnamefont
  {Vainio}}, \bibinfo {author} {\bibfnamefont {J.}~\bibnamefont {Ahokas}},
  \bibinfo {author} {\bibfnamefont {S.}~\bibnamefont {Novotny}}, \bibinfo
  {author} {\bibfnamefont {S.}~\bibnamefont {Sheludyakov}}, \bibinfo {author}
  {\bibfnamefont {D.}~\bibnamefont {Zvezdov}}, \bibinfo {author} {\bibfnamefont
  {K.-A.}\ \bibnamefont {Suominen}}, \ and\ \bibinfo {author} {\bibfnamefont
  {S.}~\bibnamefont {Vasiliev}},\ }\href {\doibase
  10.1103/PhysRevLett.108.185304} {\bibfield  {journal} {\bibinfo  {journal}
  {Physical Review Letters}\ }\textbf {\bibinfo {volume} {108}},\ \bibinfo
  {pages} {185304} (\bibinfo {year} {2012}{\natexlab{a}})}\BibitemShut
  {NoStop}%
\bibitem [{\citenamefont {Bashkin}(1981)}]{Bashkin_1981}%
  \BibitemOpen
  \bibfield  {author} {\bibinfo {author} {\bibfnamefont {E.~P.}\ \bibnamefont
  {Bashkin}},\ }\href@noop {} {\bibfield  {journal} {\bibinfo  {journal} {JETP
  Letters}\ }\textbf {\bibinfo {volume} {33}},\ \bibinfo {pages} {8} (\bibinfo
  {year} {1981})}\BibitemShut {NoStop}%
\bibitem [{\citenamefont {Lhuillier}\ and\ \citenamefont
  {Lalo{\"{e}}}(1982{\natexlab{a}})}]{Lhuillier_1982_I}%
  \BibitemOpen
  \bibfield  {author} {\bibinfo {author} {\bibfnamefont {C.}~\bibnamefont
  {Lhuillier}}\ and\ \bibinfo {author} {\bibfnamefont {F.}~\bibnamefont
  {Lalo{\"{e}}}},\ }\href@noop {} {\bibfield  {journal} {\bibinfo  {journal}
  {Le Journal de Physique}\ }\textbf {\bibinfo {volume} {43}},\ \bibinfo
  {pages} {197} (\bibinfo {year} {1982}{\natexlab{a}})}\BibitemShut {NoStop}%
\bibitem [{\citenamefont {Bouchaud}\ and\ \citenamefont
  {Lhuillier}(1985)}]{Bouchaud_1985}%
  \BibitemOpen
  \bibfield  {author} {\bibinfo {author} {\bibfnamefont {J.-P.}\ \bibnamefont
  {Bouchaud}}\ and\ \bibinfo {author} {\bibfnamefont {C.}~\bibnamefont
  {Lhuillier}},\ }\href@noop {} {\bibfield  {journal} {\bibinfo  {journal} {Le
  Journal de Physique}\ }\textbf {\bibinfo {volume} {46}},\ \bibinfo {pages}
  {1781} (\bibinfo {year} {1985})}\BibitemShut {NoStop}%
\bibitem [{\citenamefont {Lhuillier}\ and\ \citenamefont
  {Lalo{\"{e}}}(1982{\natexlab{b}})}]{Lhuillier_1982_II}%
  \BibitemOpen
  \bibfield  {author} {\bibinfo {author} {\bibfnamefont {C.}~\bibnamefont
  {Lhuillier}}\ and\ \bibinfo {author} {\bibfnamefont {F.}~\bibnamefont
  {Lalo{\"{e}}}},\ }\href@noop {} {\bibfield  {journal} {\bibinfo  {journal}
  {Le Journal de Physique}\ }\textbf {\bibinfo {volume} {43}},\ \bibinfo
  {pages} {225} (\bibinfo {year} {1982}{\natexlab{b}})}\BibitemShut {NoStop}%
\bibitem [{\citenamefont {Silvera}\ and\ \citenamefont
  {Walraven}(1986)}]{BlueBible}%
  \BibitemOpen
  \bibfield  {author} {\bibinfo {author} {\bibfnamefont {I.~F.}\ \bibnamefont
  {Silvera}}\ and\ \bibinfo {author} {\bibfnamefont {J.~T.~M.}\ \bibnamefont
  {Walraven}},\ }\href@noop {} {\emph {\bibinfo {title} {{Prog. in Low Temp.
  Phys.}}}}\ (\bibinfo  {publisher} {ed. by D. F. Brewer, North-Holland,
  Amsterdam, Vol. X, p. 139},\ \bibinfo {year} {1986})\ p.\ \bibinfo {pages}
  {139}\BibitemShut {NoStop}%
\bibitem [{\citenamefont {Vainio}\ \emph
  {et~al.}(2012{\natexlab{b}})\citenamefont {Vainio}, \citenamefont {Ahokas},
  \citenamefont {Novotny}, \citenamefont {Sheludyakov}, \citenamefont
  {Zvezdov}, \citenamefont {Suominen},\ and\ \citenamefont
  {Vasiliev}}]{Vainio_2012}%
  \BibitemOpen
  \bibfield  {author} {\bibinfo {author} {\bibfnamefont {O.}~\bibnamefont
  {Vainio}}, \bibinfo {author} {\bibfnamefont {J.}~\bibnamefont {Ahokas}},
  \bibinfo {author} {\bibfnamefont {S.}~\bibnamefont {Novotny}}, \bibinfo
  {author} {\bibfnamefont {S.}~\bibnamefont {Sheludyakov}}, \bibinfo {author}
  {\bibfnamefont {D.}~\bibnamefont {Zvezdov}}, \bibinfo {author} {\bibfnamefont
  {K.-A.}\ \bibnamefont {Suominen}}, \ and\ \bibinfo {author} {\bibfnamefont
  {S.}~\bibnamefont {Vasiliev}},\ }\href@noop {} {\bibfield  {journal}
  {\bibinfo  {journal} {Physical Review Letters}\ } (\bibinfo {year}
  {2012}{\natexlab{b}})}\BibitemShut {NoStop}%
\bibitem [{\citenamefont {Vasilyev}\ \emph {et~al.}(2004)\citenamefont
  {Vasilyev}, \citenamefont {J{\"{a}}rvinen}, \citenamefont {Tjukanoff},
  \citenamefont {Kharitonov},\ and\ \citenamefont {Jaakkola}}]{Vasilyev_2004}%
  \BibitemOpen
  \bibfield  {author} {\bibinfo {author} {\bibfnamefont {S.}~\bibnamefont
  {Vasilyev}}, \bibinfo {author} {\bibfnamefont {J.}~\bibnamefont
  {J{\"{a}}rvinen}}, \bibinfo {author} {\bibfnamefont {E.}~\bibnamefont
  {Tjukanoff}}, \bibinfo {author} {\bibfnamefont {A.}~\bibnamefont
  {Kharitonov}}, \ and\ \bibinfo {author} {\bibfnamefont {S.}~\bibnamefont
  {Jaakkola}},\ }\href@noop {} {\bibfield  {journal} {\bibinfo  {journal}
  {Review of Scientific Instruments}\ }\textbf {\bibinfo {volume} {75}},\
  \bibinfo {pages} {94} (\bibinfo {year} {2004})}\BibitemShut {NoStop}%
\bibitem [{\citenamefont {Tommila}\ \emph {et~al.}(1987)\citenamefont
  {Tommila}, \citenamefont {Tjukanov}, \citenamefont {Krusius},\ and\
  \citenamefont {Jaakkola}}]{Tommila1987}%
  \BibitemOpen
  \bibfield  {author} {\bibinfo {author} {\bibfnamefont {T.}~\bibnamefont
  {Tommila}}, \bibinfo {author} {\bibfnamefont {E.}~\bibnamefont {Tjukanov}},
  \bibinfo {author} {\bibfnamefont {M.}~\bibnamefont {Krusius}}, \ and\
  \bibinfo {author} {\bibfnamefont {S.}~\bibnamefont {Jaakkola}},\ }\href
  {http://prb.aps.org/abstract/PRB/v36/i13/p6837_1} {\bibfield  {journal}
  {\bibinfo  {journal} {Physical Review B}\ }\textbf {\bibinfo {volume} {36}},\
  \bibinfo {pages} {6837} (\bibinfo {year} {1987})}\BibitemShut {NoStop}%
\bibitem [{\citenamefont {Borovik-Romanov}\ \emph {et~al.}(1984)\citenamefont
  {Borovik-Romanov}, \citenamefont {Bunkov}, \citenamefont {Dmitriev},\ and\
  \citenamefont {Mukharskii}}]{Borovik-Romanov_1984}%
  \BibitemOpen
  \bibfield  {author} {\bibinfo {author} {\bibfnamefont {A.~S.}\ \bibnamefont
  {Borovik-Romanov}}, \bibinfo {author} {\bibfnamefont {Y.~M.}\ \bibnamefont
  {Bunkov}}, \bibinfo {author} {\bibfnamefont {V.~V.}\ \bibnamefont
  {Dmitriev}}, \ and\ \bibinfo {author} {\bibfnamefont {Y.~M.}\ \bibnamefont
  {Mukharskii}},\ }\href@noop {} {\bibfield  {journal} {\bibinfo  {journal}
  {JETP Letters}\ }\textbf {\bibinfo {volume} {40}},\ \bibinfo {pages} {1033}
  (\bibinfo {year} {1984})}\BibitemShut {NoStop}%
\bibitem [{\citenamefont {Bunkov}\ and\ \citenamefont
  {Volovik}(2010)}]{Bunkov_2010}%
  \BibitemOpen
  \bibfield  {author} {\bibinfo {author} {\bibfnamefont {Y.~M.}\ \bibnamefont
  {Bunkov}}\ and\ \bibinfo {author} {\bibfnamefont {G.~E.}\ \bibnamefont
  {Volovik}},\ }\href@noop {} {\bibfield  {journal} {\bibinfo  {journal}
  {Journal of Physics: Condensed Matter}\ }\textbf {\bibinfo {volume} {22}},\
  \bibinfo {pages} {164210} (\bibinfo {year} {2010})}\BibitemShut {NoStop}%
\bibitem [{\citenamefont {Bugrij}\ and\ \citenamefont
  {Loktev}(2008)}]{Bugrij_2008}%
  \BibitemOpen
  \bibfield  {author} {\bibinfo {author} {\bibfnamefont {A.~I.}\ \bibnamefont
  {Bugrij}}\ and\ \bibinfo {author} {\bibfnamefont {V.~M.}\ \bibnamefont
  {Loktev}},\ }\href@noop {} {\bibfield  {journal} {\bibinfo  {journal} {Low
  Temperature Physics}\ }\textbf {\bibinfo {volume} {34}},\ \bibinfo {pages}
  {992} (\bibinfo {year} {2008})}\BibitemShut {NoStop}%
\bibitem [{\citenamefont {Safonov}(2013)}]{Safonov_V_2013}%
  \BibitemOpen
  \bibfield  {author} {\bibinfo {author} {\bibfnamefont {V.~L.}\ \bibnamefont
  {Safonov}},\ }\href@noop {} {\emph {\bibinfo {title} {{Nonequillibrium
  Magnons}}}}\ (\bibinfo  {publisher} {Wiley-VCH Verlag GmbH\&Co.},\ \bibinfo
  {year} {2013})\ pp.\ \bibinfo {pages} {115--118}\BibitemShut {NoStop}%
\bibitem [{\citenamefont {L{\'{e}}vy}\ and\ \citenamefont
  {Ruckenstein}(1984)}]{Levy_1984}%
  \BibitemOpen
  \bibfield  {author} {\bibinfo {author} {\bibfnamefont {L.~P.}\ \bibnamefont
  {L{\'{e}}vy}}\ and\ \bibinfo {author} {\bibfnamefont {A.~E.}\ \bibnamefont
  {Ruckenstein}},\ }\href@noop {} {\bibfield  {journal} {\bibinfo  {journal}
  {Physical Review Letters}\ }\textbf {\bibinfo {volume} {52}},\ \bibinfo
  {pages} {1512} (\bibinfo {year} {1984})}\BibitemShut {NoStop}%
\end{thebibliography}%

\end{document}